\documentclass[a4paper,11pt]{article}
\pdfoutput=1 

\usepackage{jheppub} 

\usepackage[T1]{fontenc} 
\usepackage{mathrsfs}
\usepackage{latexsym,bm}
\usepackage{graphicx}
\usepackage{indentfirst}
\usepackage{slashed}
\usepackage{amssymb}
\usepackage{amsmath}
\allowdisplaybreaks[4] 
\usepackage{bbm}
\usepackage{color}
\usepackage[dvipsnames]{xcolor}
\usepackage{epsfig}
\usepackage[titletoc]{appendix}
\usepackage{multirow}%
\usepackage{rotating}
\usepackage{epstopdf}
\usepackage{extarrows}
\usepackage{tabularx}
\usepackage{soul} 
\usepackage{xcolor}
\usepackage{lipsum}

\usepackage[utf8]{inputenc} 

\newcommand{\be}{\begin{equation}} \newcommand{\ee}{\end{equation}}
\newcommand{\ba}{\begin{array}{c}} \newcommand{\ea}{\end{array}}
\newcommand{\bea}{\begin{eqnarray}} \newcommand{\eea}{\end{eqnarray}}

\newcommand{\rd}{{\rm d}}

\newcommand{\vecA}{\widehat{A}}
\newcommand{\vecB}{\widehat{B}}
\newcommand{\vecC}{\widehat{C}}
\newcommand{\vecD}{\widehat{D}}
\newcommand{\vecT}{\widehat{T}}
\newcommand{\tdB}{\widetilde{B}}

\title{\boldmath A unified formulation of one-loop tensor integrals 
for finite volume effects}

\author[a]{Ze-Rui~Liang,}
\author[a,b,c,1]{De-Liang~Yao\note{Corresponding author.}}

\affiliation[a]{School of Physics and Electronics, Hunan University, 410082 Changsha, China}
\affiliation[b]{Hunan Provincial Key Laboratory of High-Energy Scale Physics and Applications, Hunan University, 410082 Changsha, China}
\affiliation[c]{School for Theoretical Physics, Hunan University, 410082 Changsha, China}

\emailAdd{zeruiliang@hnu.edu.cn}
\emailAdd{yaodeliang@hnu.edu.cn}

\abstract{A unified formulation of one-loop tensor integrals is proposed for systematical calculations of finite volume corrections. It is shown that decomposition of the one-loop tensor integrals into a series of tensors accompanied by tensor coefficients is feasible, if a unit space-like four vector $n^\mu$, originating from the discretization effects at finite volume, is introduced. A generic formula has been derived for numerical computations of all the involved tensor coefficients. For the vanishing external three-momenta, we also investigate the feasibility of the conventional Passarino-Veltmann reduction of the tensor integrals in a finite volume. Our formulation can be easily used to realize the automation of the calculations of finite volume corrections to any interesting quantities at one-loop level. Besides, it provides finite volume result in a unique and concise form, which is suited for, e.g., carrying out precision determination of physical observable from modern lattice QCD data.
}

\begin{document} 
\maketitle
\flushbottom

\section{Introduction}
\label{sec:intro}
Finite volume effects are one of the major sources of systematic uncertainties in modern lattice QCD computations. A primary task in lattice QCD is to evaluate the so-called finite volume corrections (FVC). Especially when Monte Carlo simulations are performed with pion mass close to its physical value, the size of FVC becomes large. Therefore, proper treatment of finite-size effects has been of crucial importance in the precise extraction of physical results from Monte Carlo data. Even though lattice data for a quantity of interest are not available for the time being, it is still important to estimate the size of FVC so as to choose an appropriate lattice, in which the unwanted finite volume artifacts are suppressed. Finite volume effects are also significant in their own right and imply a few interesting physical origins. As pointed out by L\"uscher in Ref.~\cite{Luscher:1985dn}, for a stable particle, finite-size effects may result in a mass shift, which can be physically interpreted as the existence of a cloud of virtual light particles around it. 

In practice, a systematical and popular tool to evaluate FVC is the chiral perturbation theory (ChPT)~\cite{Weinberg:1978kz,Gasser:1983yg,Gasser:1984gg} at finite volume. It was proposed, by Gasser and Leutwyler in Refs.~\cite{Gasser:1986vb,Gasser:1987zq}, that it is feasible to do FVC calculations in the so-called $p$-regime by using the chiral effective Lagrangians constructed in the infinite volume. For a brief review, see e.g. Refs.~\cite{Meissner:1993ah,Colangelo:2004sc}. The merit is that the obtained FVC results depend both on the finite spatial extent $L$ and on the pion mass $M_\pi$, and hence can be utilized to perform chiral extrapolation ($m_{u/d}\to m_{u/d}^{\rm phys.}$) and also themodynamical extrapolation   ($L\to\infty$) of lattice data. Based on ChPT, there accumulate a multitude of works concerning finite volume calculations, e.g.,  masses~\cite{Beane:2004tw,Alvarez-Ruso:2013fza,Yao:2018ifh,Severt:2020jzc}, decay constants~\cite{Becirevic:2003wk,Geng:2014efa}, nucleon electric dipole moments~\cite{Akan:2014yha}, scalar form factor in $K_{\ell3}$ semi-leptonic decay~\cite{Ghorbani:2010bh}, etc.. Very recently, FVC to forward Compton scattering off the nucleon has been derived in the framework of manifestly covariant baryon ChPT~\cite{delaParra:2020rct}. In particular, two-loop sunset integrals in the finite volume are computed for pion mass and decay constants in Refs.~\cite{Colangelo:2006mp,Bijnens:2013doa,Bijnens:2014dea}. A package for numerical computation of the one-loop tadpole and two-loop sunset integrals is realized in Ref.~\cite{Bijnens:2014gsa}.

On the other hand, the L\"uscher formula~\cite{Luscher:1985dn} provides an alternative approach to calculate FVC to masses. Its application to the study of the FVC to the masses of pion, nucleon and heavy mesons can be found in Refs.~\cite{Colangelo:2003hf,Colangelo:2005gd,Colangelo:2010ba}. In this method, the FVC of the masses are related to the forward scattering amplitudes which, in the low-energy region, are obtainable from ChPT. The advantage of the L\"uscher formula lies in that the leading exponential term $\exp(-M_\pi L)$ in the finite-volume dependence can be easily estimated~\cite{Luscher:1985dn}. It was shown in Ref.~\cite{Colangelo:2005gd} that an input of a tree-level ChPT scattering amplitude can reproduce the one-loop FVC result evaluated in ChPT at finite volume. Nevertheless, the L\"uscher approach fails in generating exponential terms beyond leading order. A resummed version of the L\"uscher formula was advocated in Ref.~\cite{Colangelo:2005gd} to restore the subleading exponential corrections. Besides, a L\"uscher-formula-like asymptotic expression was derived for the study of decay constant~\cite{Colangelo:2004xr}. In a word, the feasibility of the L\"uscher formula approach is rather limited. Hence, in many cases, one is often suggested to prefer the method of ChPT at finite volume, as introduced in the preceding paragraph.

However, the calculations of FVC within the framework of ChPT at finite volume are tedious. Complexity has already taken place in the one-loop analyses, see e.g. Refs.~\cite{Becirevic:2003wk,Beane:2004tw,Ghorbani:2010bh,Alvarez-Ruso:2013fza,Geng:2014efa,Akan:2014yha,Yao:2018ifh,delaParra:2020rct}. As far as we know, automation of the one-loop calculations of FVC is still unavailable. Even worse, expressions of the results for a given quantity might be different in form. Namely, a shift of the integration momentum in the integral or an inversion of the direction of the momentum in the loop graph would lead to different forms of analytical results, although the corresponding numerical result keeps the same. It should be mentioned that the analytical outcomes are usually expressed in terms of Bessel functions. In principle, one is able to translate them into a unique form by employing the recurrence relations of the Bessel functions, but the procedure is often intricate. The above-mentioned issues actually are due to the lack of a unified description of the one-loop tensor integrals in a finite volume. In this work, we intend to fill this gap.

In the infinite volume, it has been standard to express  one-loop tensor integrals in terms of Lorentz-covariant structures multiplied by Lorentz tensors $L^{\mu_1\cdots\mu_P}$ of rank $P$, which are composed of metric tensors and external momenta; see e.g. Refs.~\cite{Passarino:1978jh,Denner:1991kt,Denner:2002ii,Denner:2005nn} for details. Unfortunately, the decomposition does not hold true for the integrals in a finite volume any longer. The main difficulty is owing to the un-welcome occurrence of an extra vector $l^\mu=(0,\mathbf{n}L)\equiv n^\mu L$ with $\mathbf{n}\in \mathbb{Z}^3$, in addition to the external momenta and metric tensor. This vector is introduced by the boundary conditions of the finite volume $V=L^3$ and, hereafter, $n^\mu$ is called the unit space-like four vector for clarity. Thus, on top of terms proportional to the Lorentz tensors $L^{\mu_1\cdots\mu_P}$ in the infinite case, the results of FVC contain pieces related to $n^\mu$ as well. To address this issue, we generalize the Lorentz tensor, by incorporating the unit space-like four vector $n^\mu$, to $\widetilde{L}^{\mu_1\cdots\mu_P}$ as shown in Eq.~\eqref{eq:p.rank.tensor}. With the help of the new rank-$P$ tensor, any one-loop tensor integrals for FVC can be decomposed uniquely. More importantly, we find that a compact formula can be achieved for the tensor coefficients after a direct calculation using Poisson summation formula and Feynman parameterization techniques. In fact, this formula in turn justifies the validity of the decomposition method we have proposed here.

The reduction of tensor integrals has a long history~\cite{Passarino:1978jh,vanNeerven:1983vr}, but still continues to intrigue quite a few studies nowadays~\cite{Diakonidis:2009fx,Lyubovitskij:2021ges,Feng:2022uqp}. Generally speaking, the symmetry of Lorentz invariance in the infinite volume allows one to reduce all tensor coefficients to the basic scalar integrals~\cite{tHooft:1978jhc}. Various algorithms have been invented to either make the process of reduction faster or to tackle special cases concerning singularities, see e.g. Ref.~\cite{Lyubovitskij:2021ges} and the references therein. Here, it would be also interesting to investigate the reduction of the tensor integrals at finite volume. Unlike the infinite case, problem arises because the Lorentz invariance is explicitly broken in a finite volume. Such a problem can be avoided in the frame defined by the conditions $n\cdot p_i=0$, where $p_i$ are external momenta and $n$ is the unit space-like four vector. For definiteness, this frame is referred to as \lq\lq center-of-mass" (CM) frame throughout this work.\footnote{
Note that, since $n^\mu=(0,{\bf n})$ is arbitrary, the constraint $n\cdot p_i=0$ leads to ${\bf p}_i=0$ for $i=1,\cdots, N-1$, where $N$ is the number of denominator factors defined in Eq.~\eqref{eq:tensor.int.def0}. Although the validity of the obtained results in Section~\ref{sec:reduction} is unaffected, the realistic application of, e.g., reduction of the FVC tensor coefficients, is actually rather limited, especially for $N\geq 3$.} In the CM frame, two findings can be achieved. One is that the tensor coefficients can be indeed reduced recursively and two universal reduction formulae are obtained. The other is that, in addition to the scalar integrals, the tensor coefficients proportional to the metric tensor should be incorporated into the basis. This is a new feature of the one-loop tensor integrals for FVC, compared to the ones in the infinite volume.

Our unified formulation of one-loop tensor integrals for FVC can be easily implemented in, e.g. FeynCalc~\cite{Mertig:1990an,Shtabovenko:2016sxi,Shtabovenko:2020gxv}, in future, so as to realize the automation of the calculations of FVC to any interesting quantities at one-loop level. As a first application, our formulation is used to compute FVC to the nucleon mass. The obtained results are concise. The relevant result in previous literature can be reproduced. We also check that the results with and without the Passarino-Veltman (PV) reduction are exactly the same, indicating the correctness of PV reduction numerically.

The layout of this manuscript is described as follows. In Section~\ref{sec:decomposition}, the decomposition of loop integrals for FVC is discussed in detail, and a general formula for numerical computation of FVC tensor coefficients is derived. Section~\ref{sec:reduction} comprises the PV reduction of the tensor coefficients in the CM frame. An example is shown in Section~\ref{sec:application} to illustrate the application of our unified formulation. Section~ \ref{sec:summary} contains summary and outlook. In Appendix~\ref{app.formulae}, some useful formulae we used are listed. The properties of the auxiliary tensor $h_{\mu\nu}$ are briefly presented in Appendix~\ref{app.h.tensor}. A detailed mathematical proof of Eq.~\eqref{eq:tensortoscalar} is provided in Appendix~\ref{app.proof}. For the sake of easy reference, some results of one-loop tensor integrals with the number of internal propagators less than and equal to four are collected in Appendix~\ref{app.coeff}. Correspondingly, explicit expressions regarding the decomposition of tensor integrals in the CM frame are shown in Appendix~\ref{app.coeff.cm}.
Finally, Appendix~\ref{app.se} comprises explicit expressions of the one-loop self-energies of the nucleon.

\section{Decomposition of one-loop tensor integrals}\label{sec:decomposition}

As already mentioned in the Introduction, ChPT allows one to perform systematic evaluations of physical quantities of interest both in the infinite space and at finite volume. In the infinite volume, high-order chiral calculations have already became standard and efficient, benefiting greatly from the well-prepared loop integrals defined in literature~\cite{tHooft:1978jhc,Passarino:1978jh,Denner:2005nn} and in some relevant packages~\cite{Mertig:1990an,Shtabovenko:2016sxi,Shtabovenko:2020gxv,Hahn:1998yk}. On the contrary, studies done in a finite volume always suffer from tedious computations of loop integrals from the very beginning. To tackle this issue, in this section, we are going to provide a general description of one-loop tensor integrals at finite volume. Note that this can be achieved, equivalently, by studying the finite-volume shifts of the loop integrals. In what follows, one-loop tensor integrals are defined for FVC, and pertinent tensor decomposition of those integrals is detailed. And then, we 
will derive a uniform formula for the tensor coefficients appearing in the decomposition.

\subsection{Definition of loop integrals for FVC}
The general form of one-loop $N$-point rank-$P$ tensor integrals can be written as 
\begin{align}\label{eq:tensor.int.def0}
{T}^{N,\mu_1,\cdots,\mu_P}=
\frac{1}{i}
\int \frac{\rd^dk}{(2\pi)^d}\frac{k^{\mu_1}\cdots k^{\mu_P}}{D_1D_2\cdots D_N}
\ ,
\end{align}
where the denominator factors are defined by
\begin{align}
D_j=[(k+p_{j-1})^2-m_j^2+i0^+]\ ,\quad j=1,\cdots,N \ ,
\end{align}
with $p_0=0$ and $i0^+$ being an infinitesimal imaginary part. Note that the notation of Ref.~\cite{tHooft:1978jhc} is adopted to denote the individual $N$-point integrals: $T^1=A$, $T^2=B$, $T^3=C$, $T^4=D$, $T^5=E$, $\cdots$. Furthermore, the case of $P=0$ corresponds to the scalar integral, which is conventionally denoted by $T^N_0$, e.g., $T^1_0=A_0$, $T^2_0=B_0$, etc.. For brevity, the arguments of ${T}^{N,\mu_1,\cdots,\mu_P}$ are suppressed.  In a cubic box of volume $V=L^3$, the integration over the internal momentum $\mathbf{k}$ should be replaced by a sum over all allowed discrete momenta. Specifically, for an arbitrary function $F(\mathbf{k})$, one has
\begin{align}
\int\frac{\rd\mathbf{k}}{(2\pi)^3} F(\mathbf{k})\longrightarrow \frac{1}{L^3}\sum_{\mathbf{n}} F(\mathbf{k}_n)\ ,
\end{align}
where $\mathbf{k}_n={2\pi\mathbf{n}}/{L}$ and $\mathbf{n}\equiv (n_1,n_2,n_3)$ with $n_i\in \mathbb{Z}$ ($i=1,2,3$)\footnote{Here the discretized momenta arise from the periodic boundary conditions. For discussions on the momenta constrained by twisted boundary conditions, see e.g. Ref.~\cite{Colangelo:2016wgs}.}. 
Therefore, the tensor integrals in Eq.~\eqref{eq:tensor.int.def0} at finite volume are given by
\begin{align}
{T}^{N,\mu_1,\cdots,\mu_P}_V=\frac{1}{i}\left(\frac{1}{L^3}\sum_{\mathbf{n}}\int\frac{\rd k^0}{2\pi}\right)
\frac{k^{\mu_1}\cdots k^{\mu_P}}{D_1D_2\cdots D_N}\ 
\equiv 
\frac{1}{i}
\int_V \frac{\rd^dk}{(2\pi)^d}\frac{k^{\mu_1}\cdots k^{\mu_P}}{D_1D_2\cdots D_N}\ .
\end{align}
Here the time extent is left infinite, due to the assumption that it is much larger than the spatial length $L$ of the cubic box. Actually, that is the usual situation encountered in lattice QCD simulations. The finite-volume sums in the above equation can be done by making use of Poisson summation formula~\cite{stein2011fourier}, i.e.
\begin{align}
\frac{1}{L^3}\sum_{\mathbf{n}}F(\mathbf{k}_n)=\sum_{\mathbf{n}}\int\frac{\rd \mathbf{k}}{(2\pi)^3}e^{i\mathbf{l}_{k}\cdot\mathbf{k}}F(\mathbf{k}_n)\ .
\end{align}
Then, the finite-volume tensor integrals are transformed into
\begin{align}\label{eq:tensorint2}
{T}^{N,\mu_1,\cdots,\mu_P}_V=\sum_{\mathbf{n}}\frac{1}{i}
\int\frac{\rd^dk}{(2\pi)^d}e^{-i{l}_{k}\cdot {k}}
\frac{k^{\mu_1}\cdots k^{\mu_P}}{D_1D_2\cdots D_N}\ ,
\end{align}
where a four vector $l_k^\mu\equiv (0,\mathbf{n}L)$ has been introduced. One can notice that the term with $l_k=0$, i.e. $\mathbf{n}=0$, in the sum represents the infinite-volume contribution, since it coincides with Eq.~\eqref{eq:tensor.int.def0}. 

The difference between ${T}_V^{N,\mu_1,\cdots,\mu_P}$ and ${T}^{N,\mu_1,\cdots,\mu_P}$ defines the so-called FVC, i.e.
\begin{align}
\widetilde{T}^{N,\mu_1,\cdots,\mu_P}\equiv T^{N,\mu_1,\cdots,\mu_P}_{V}-T^{N,\mu_1,\cdots,\mu_P}\ .
\end{align}
Hereafter, we use the notation $\widetilde{T}$ to represent the tensor integrals defined for the calculation of FVC. Inserting Eq.~\eqref{eq:tensor.int.def0} and Eq.~\eqref{eq:tensorint2} to the above equation, the tensor integrals for FVC are given by
\begin{align}\label{eq:FVC.tensor.int.newform}
\widetilde{T}^{N,\mu_1,\cdots,\mu_P}=\sum_{\mathbf{n}\neq 0}\frac{1}{i}
\int\frac{\rd^dk}{(2\pi)^d}e^{-i{l}_{k}\cdot {k}}
\frac{k^{\mu_1}\cdots k^{\mu_P}}{D_1D_2\cdots D_N}\ .
\end{align}
Note that FVC are ultraviolet (UV) convergent and one can set $d=4$. Nevertheless, the symbol $d$ is retained explicitly throughout this paper.
 It can be inferred that, due to the emergence of the exponential piece $e^{-il_k\cdot k}$, the results after integration over $k$ shall depend on the four vector $l_k$ in addition to the external four momenta $p_j$ ($j=1,\cdots,N-1$). Consequently, the Lorentz indices $\mu_1,\cdots,\mu_P$ are distributed over the $p_j$'s, $l_k$ and the metric tensor, which will be discussed in the next subsection.

\subsection{Decomposition of the FVC tensor integrals\label{sec.decom}}
In the infinite volume, the tensor integrals can be decomposed into Lorentz-covariant structures, as illustrated, e.g., in Refs.~\cite{Denner:2002ii,Denner:2005nn}. Similar procedure of tensor decomposition can be applied to the FVC tensor integrals. 

To that end, a typical rank-$P$ tensor is introduced,
\begin{align}\label{eq:p.rank.tensor}
\widetilde{L}^{\mu_1\cdots\mu_P}=\{\underbrace{g\cdots g}_s\, p\cdots p\,\underbrace{n\cdots n}_r\}^{\mu_1\cdots\mu_P}_{i_{2s+1}\cdots i_{P-2s-r}}\ ,
\end{align}
where $n^\mu=(0,\mathbf{n})$ is the unit space-like four vector , which is related to $l_k$ via $l_k^\mu=n^\mu L$. The rank-$P$ tensor comprises $s$ metric tensors, $r$ unit space-like vectors  and $P-2s-r$ momenta $p_{i_{2s+1}},\cdots, p_{i_{P-2s-r}}$, with $i_{2s+1},\cdots, i_{P-2s-r}\in\{1,\cdots,N-1\}$. Furthermore, the curly braces denote that the Lorentz indices are assigned as follows. 
First, $2s$ out of $P$ Lorentz indices are distributed over the metric tensors and any pair of them are symmetrical. 
Second, the $n$-vectors occupy $r$ Lorentz indices from the remaining ones. Third, the rest Lorentz indices are assigned to the momenta $p_{i_{2s+1}}^{\mu_{2s+1}}\cdots p_{i_{P-2s-r}}^{\mu_{P-2s-r}}$. Only one representative out of the $P-2s-r$ permutations of the indices $i_j$ is kept.
As a result, the number of terms appearing in Eq.~\eqref{eq:p.rank.tensor} is 
\begin{align}
    \frac{C_P^2C_{P-2}^2\cdots C_{P-2s+2}^2}{s!}\times C_{P-2s}^r=\frac{P!}{2^s\,s!\,r!\,(P-2s-r)!}
    \ .
\end{align}
Some instructive examples are given below:
\begin{align}
&\{ p p \cdots p \}^{\mu_1 \mu_2 \cdots \mu_P}_{i_1 i_2 \cdots i_P}
=
p^{\mu_1}_{i_1} p^{\mu_2}_{i_2} \cdots p^{\mu_P}_{i_P} \ , \\
&\{ p n \}^{\mu_1\mu_2}_{i_1}
=
p^{\mu_1}_{i_1} n^{\mu_2}
+
n^{\mu_1} p^{\mu_2}_{i_1}
\ , \\
&\{ p p n \}^{\mu_1\mu_2\mu_3}_{i_1i_2}
=
p^{\mu_1}_{i_1} p^{\mu_2}_{i_2} n^{\mu_3}
+
p^{\mu_1}_{i_1} n^{\mu_2} p^{\mu_3}_{i_2}
+
n^{\mu_1} p^{\mu_2}_{i_1} p^{\mu_3}_{i_2} 
\ , \\
&\{ p n n \}^{\mu_1\mu_2\mu_3}_{i_1}
=
p^{\mu_1}_{i_1} n^{\mu_2} n^{\mu_3}
+
n^{\mu_1} p^{\mu_2}_{i_1} n^{\mu_3} 
+
n^{\mu_1} n^{\mu_2} p^{\mu_3}_{i_1}
\ , \\
&\{ g n \}^{\mu_1\mu_2\mu_3}
= g^{\mu_1\mu_2} n^{\mu_3}
+ g^{\mu_1\mu_3} n^{\mu_2}
+ g^{\mu_2\mu_3} n^{\mu_1}
\ , \\
&\{ g p n\}^{\mu_1\mu_2\mu_3\mu_4}_{i_1}
=
g^{\mu_1\mu_2} 
(p^{\mu_3}_{i_1} n^{\mu_4}+n^{\mu_3} p^{\mu_4}_{i_1})
+
g^{\mu_1\mu_3} 
(p^{\mu_2}_{i_1} n^{\mu_4}+n^{\mu_2} p^{\mu_4}_{i_1})
\notag \\
&
\qquad \qquad \quad \quad 
+
g^{\mu_1\mu_4} 
(p^{\mu_2}_{i_1} n^{\mu_3}+n^{\mu_2} p^{\mu_3}_{i_1})
+
g^{\mu_2\mu_3} 
(p^{\mu_1}_{i_1} n^{\mu_4}+n^{\mu_1}p^{\mu_4}_{i_1})
\notag \\
&
\qquad \qquad \quad \quad 
+
g^{\mu_2\mu_4} (p^{\mu_1}_{i_1} n^{\mu_3}
+n^{\mu_1} p^{\mu_3}_{i_1})
+
g^{\mu_3\mu_4} (p^{\mu_1}_{i_1} n^{\mu_2}+n^{\mu_1} p^{\mu_2}_{i_1})
\ , \\
&\{ g g \}^{\mu_1\mu_2\mu_3\mu_4}
=
g^{\mu_1\mu_2} g^{\mu_3\mu_4}
+
g^{\mu_1\mu_3} g^{\mu_2\mu_4}
+
g^{\mu_1\mu_4} g^{\mu_2\mu_3}
\ .
\end{align}

With the help of the rank-$P$ tensor, one can decompose the one-loop tensor integrals into the form as
\begin{align}
\label{eq:FVC.int.def-1}
\widetilde{T}^{N,\mu_1\mu_2\cdots\mu_P}
=\sum_{\mathbf{n}\neq 0} \vecT^{N,\mu_1\mu_2\cdots\mu_P}\ ,
\end{align}
with 
\begin{align}
\vecT^{N,\mu_1\mu_2\cdots\mu_P}&=
\sum^{[{P}/{2}]}_{s=0}
\sum^{P-2s}_{r=0}
\sum^{N-1}_{\begin{subarray}{c}
i_{2s+1}=1, \\
\cdots \\
i_{P-2s-r}=1
\end{subarray}}
\big\{
\underbrace{g\cdots g}_{s}
p\cdots p 
\underbrace{n\cdots n}_{r}
\big\}^{\mu_1\cdots \mu_P}_{i_{2s+1}\cdots i_{P-2s-r}}
\vecT^N_{\underbrace{\scriptstyle 0\cdots 0}_{2s}i_{2s+1}\cdots i_{P-2s-r}
\underbrace{\scriptstyle N\cdots N}_{r}}
\ .\label{eq:FVC.int.def}
\end{align}
Here $[{P}/{2}]$ is the floor function, which maps a real number to the largest integer less than or equal to ${P}/{2}$. The hat over $\vecT$ indicates the dependence on the vector $\mathbf{n}$. It should be mentioned that the tensor coefficient $\vecT^N_{\scriptstyle 0\cdots 0i_{2s+1}\cdots i_{P-2s-r}N\cdots N}$ is invariant with respect to permutation of the subscripts $i_j$'s.  Thus, in practice it is usual to choose a representative with the $i_j$ indices arranged in ascending order. For instance, $\widehat{C}_{001233}=\widehat{C}_{002133}$ and we choose to use $\widehat{C}_{001233}$ as the representative. It is worth pointing out that there are no counterparts in the infinite volume for those tensor coefficients with subscripts \lq\lq $N$" in Eq.~\eqref{eq:FVC.int.def}.

For future reference, decomposition of the FVC tensor integrals up to rank 5 are explicitly listed below.
\begin{align}
\vecT^{N,\mu}
&=
\sum^{N-1}_{i=1}
p^\mu_i 
\vecT^N_i
+
n^\mu 
\vecT^N_N 
\ , \\
\vecT^{N,\mu\nu}
&=
g^{\mu\nu} 
\vecT^N_{00}
+
\sum^{N-1}_{i,j=1}
p^\mu_{i} p^\nu_{j} 
\vecT^N_{i j}
+
\sum^{N-1}_{i=1}
\{ p n \}^{\mu\nu}_{i}
\vecT^N_{i N}
+
n^\mu n^\nu 
\vecT^N_{NN}
\ , \\
\vecT^{N,\mu\nu\rho}
&=
\sum^{N-1}_{i=1}
\{ g p \}^{\mu\nu\rho}_{i}
\vecT^N_{00 i}
+
\{ g n \}^{\mu\nu\rho} 
\vecT^N_{00N}
+
\sum^{N-1}_{i, j, k=1}
p^\mu_{i} p^\nu_{j} p^\rho_{k}
\vecT^{N}_{i j k}
\notag \\
&
+
\sum^{N-1}_{i, j=1}
\{ p p n \}^{\mu\nu\rho}_{ij} 
\vecT^N_{i j N}
+
\sum^{N-1}_{i=1}
\{ p n n \}^{\mu\nu\rho}_{i}
\vecT^N_{iNN}
+
n^\mu n^\nu n^\rho 
\vecT^N_{NNN}
\ , \\
\vecT^{N, \mu\nu\rho\sigma}
&=
\{ g g \}^{\mu\nu\rho\sigma} 
\vecT^N_{0000}
+
\sum^{N-1}_{i,j=1}
\{ g p p \}^{\mu\nu\rho\sigma}_{ij}
\vecT^N_{00ij}
+
\sum^{N-1}_{i=1}
\{ g p n \}^{\mu\nu\rho\sigma}_{i}
\vecT^N_{00iN}
\notag \\
&
+
\{ g n n \}^{\mu\nu\rho\sigma} 
\vecT^N_{00NN}
+
\sum^{N-1}_{i,j,k,l=1}
p^\mu_i p^\nu_j p^\rho_k p^\sigma_l 
\vecT^N_{ijkl}
+
\sum^{N-1}_{i,j,k=1}
\{ p p p n \}^{\mu\nu\rho\sigma}_{ijk}
\vecT^N_{ijkN}
\notag \\
&
+
\sum^{N-1}_{i,j=1} 
\{ p p n n \}^{\mu\nu\rho\sigma}_{ij}
\vecT^N_{ijNN}
+
\sum^{N-1}_{i=1}
\{ p n n n \}^{\mu\nu\rho\sigma}_{i}
\vecT^N_{iNNN}
+
n^\mu n^\nu n^\rho n^\sigma
\vecT^N_{NNNN}
\ , \\
\vecT^{N, \mu\nu\rho\sigma\alpha}
&=
\sum^{N-1}_{i=1}
\{ g g p \}^{\mu\nu\rho\sigma\alpha}_{i}
\vecT^N_{0000i}
+
\{ g g n \}^{\mu\nu\rho\sigma\alpha}
\vecT^N_{0000N}
+
\sum^{N-1}_{i,j,k=1}
\{ g p p p \}^{\mu\nu\rho\sigma\alpha}_{ijk}
\vecT^N_{00ijk}
\notag \\
&
+
\sum^{N-1}_{i,j=1}
\{ g p p n \}^{\mu\nu\rho\sigma\alpha}_{ij}
\vecT^N_{00ijN}
+
\sum^{N-1}_{i=1}
\{ g p n n \}^{\mu\nu\rho\sigma\alpha}_{i}
\vecT^N_{00iNN}
+
\{ g n n n \}^{\mu\nu\rho\sigma\alpha}
\vecT^N_{00NNN}
\notag \\
&
+
\sum^{N-1}_{i,j,k,l,r=1}
p^\mu_i p^\nu_j p^\rho_k p^\sigma_l p^\alpha_r
\vecT^N_{ijklr}
+
\sum^{N-1}_{i,j,k,l=1}
\{ p p p p n \}^{\mu\nu\rho\sigma\alpha}_{ijkl}
\vecT^N_{ijklN}
+
\sum^{N-1}_{i,j,k=1}
\{ p p p n n \}^{\mu\nu\rho\sigma\alpha}_{ijk}
\vecT^N_{ijkNN}
\notag \\
&
+
\sum^{N-1}_{i,j=1}
\{ p p n n n \}^{\mu\nu\rho\sigma\alpha}_{ij}
\vecT^N_{ijNNN}
+
\sum^{N-1}_{i=1}
\{ p n n n n \}^{\mu\nu\rho\sigma\alpha}_{i}
\vecT^N_{iNNNN}
+
n^\mu n^\nu n^\rho n^\sigma n^\alpha
\vecT^N_{NNNNN}
\ . \label{eq:FVC.dec}
\end{align}

Obviously, once the coefficients $\vecT^N_{\scriptstyle 0\cdots 0i_{2s+1}\cdots i_{P-2s-r}N\cdots N}$ are known, all the FVC tensor integrals are determined. In the next subsection, we will derive a compact formula suitable for the numerical computation of the tensor coefficients.

\subsection{Evaluation of the coefficients}
By the application of Feynman parameterization, the right hand side of Eq.~\eqref{eq:FVC.tensor.int.newform} can be rewritten as
\begin{align}\label{eq:Min}
\widetilde{T}^{N,\mu_1,\cdots,\mu_P}=\sum_{\mathbf{n}\neq 0}\int_0^1 \rd \mathcal{X}_N\left\{\frac{1}{i}
\int\frac{\rd^dk}{(2\pi)^d}e^{-i{l}_{k}\cdot {k}}
\frac{k^{\mu_1}\cdots k^{\mu_P}}{\left[(k+\mathcal{P}_N)^2-\mathcal{M}_N^2+i0^+\right]^N}\right\}\ ,
\end{align}
where the abbreviation $\int_0^1\rd \mathcal{X}_N\equiv\Gamma(N)\int_0^1\rd x_1\cdots\int_0^1\rd x_{N-1}x_2\cdots x_{N-1}^{N-2}$ has been used, and $x_i$ ($i=1,\cdots, N-1$) are the Feynman parameters. The $\mathcal{P}_N$ and $\mathcal{M}_N^2$ in the denominator can be obtained by using the following recursive relations
\begin{align}
&\mathcal{P}_{j+1}=x_j\mathcal{P}_j+(1-x_j)p_j   \ ,\quad \mathcal{P}_1=p_0\ ,\label{eq:mathP}\\
&\mathcal{Q}^2_{j+1}=x_j\mathcal{Q}_j^2+(1-x_j){(m_{j+1}^2-p_j^2)}\ ,\quad \mathcal{Q}_1^2={m_1^2-p_0^2}\ ,\label{eq:mathQ}\\
&\mathcal{M}_{j+1}^2=\mathcal{Q}_{j+1}^2+\mathcal{P}_{j+1}^2\ , \label{eq:mathM}
\end{align}
with $p_0=0$ and $j=1,\cdots, N-1$. 

The remaining task is to perform the momentum integration. In general, it is much easier to do the integration in Euclidean space than in Minkowski space, due to the fact that $d$-dimensional spherical coordinates can be imposed. The integral in the curly braces of Eq.~\eqref{eq:Min} is then changed to that in Euclidean space by making use of Wick rotation, which gives
\begin{align}
\bigg\{\cdots\bigg\}_E=
{(-1)^N}\int\frac{\rd^dk_E}{(2\pi)^d}e^{i{l}_{k}\cdot {k}_E}
\frac{k_E^{\mu_1}\cdots k_E^{\mu_P}}{\left[(k_E+\mathcal{P}_N^E)^{2}+\mathcal{M}_N^{E,2}\right]^N}\ .
\end{align}
Here, we define $k_E^\mu\equiv(k^0,\vec{k})$ with $k^0=ik_E^0$ and $\vec{k}=\vec{k}_E$, same for all the other involved momenta. One should keep in mind that now the corresponding metric tensor of Euclidean tensor is $\delta_{\mu\nu}={\rm diag}(1,1,1,1)$, rather than $g_{\mu\nu}$ with signature $(1,-1,-1,-1)$. 

To proceed, the Gaussian parameterization Eq.~\eqref{eq:Gaussian} is used to rewrite the denominator factors into an exponential form as 
\begin{align}
\bigg\{\cdots\bigg\}_E=
\frac{(-1)^N}{\Gamma(N)}\int\frac{\rd^dk_E}{(2\pi)^d}
\int_0^\infty\rd \lambda \lambda^{N-1}\left\{k_E^{\mu_1}\cdots k_E^{\mu_P}\right\}
e^{-\lambda\left[(k_E+\mathcal{P}_N^E)^2+\mathcal{M}_N^{E,2}\right]+i{l}_{k}\cdot {k}_E}
\ .
\end{align}
Next, we complete the square for ${k}_E$ in the exponential factor and shift it to $\bar{k}_E=k_E+\mathcal{P}^E_N-\frac{il_k}{2\lambda}$, which yields
\begin{align}
\bigg\{\cdots\bigg\}_E&=
\frac{(-1)^N}{\Gamma(N)}e^{-il_k\cdot \mathcal{P}_N^E}
\int_0^\infty\rd\lambda \lambda^{N-1}e^{-\lambda \mathcal{M}_N^{E,2}-\frac{l_k^2}{4\lambda}}\notag\\
&
\times\left\{\int\frac{\rd^d\bar{k}_E}{(2\pi)^d}\left[\bar{k}_E+\frac{il_k}{2\lambda}-\mathcal{P}_N^{ E}\right]^{\mu_1}\cdots \left[\bar{k}_E+\frac{il_k}{2\lambda}-\mathcal{P}_N^{ E}\right]^{\mu_P}e^{-\lambda \bar{k}_E^2}\right\}
\ .
\end{align}
One notices that the domain of the momentum integral is symmetric about zero, therefore the terms with odd numbers of $\bar{k}_E$'s vanish. Namely, only the terms with even numbers of $\bar{k}_E$'s survive. They can be reformulated by utilizing the following identity
\begin{align}
\bar{k}_E^{\mu_1}\cdots \bar{k}_E^{\mu_{2s}}=\frac{1}{2^s({d}/{2})_s}\{\delta\cdots \delta\}^{\mu_1\cdots\mu_{2s}}\left(\bar{k}_E^2\right)^s\ ,\notag
\end{align}
where $(d/2)_s=\Gamma(\frac{d}{2}+s)/\Gamma(\frac{d}{2})$ is the Pochhammer symbol, cf. Eq.~\eqref{eq:Pochhammer}, and the small curly braces are the same as the ones used in Eq.~\eqref{eq:p.rank.tensor}. Consequently, the integrand is a function of the magnitude of $\bar{k}_E$ and the momentum integral can be performed by employing Eq.~\eqref{eq:mom.int}. One obtains
\begin{align}
\bigg\{\cdots\bigg\}_E&=
\sum_{s=0}^{[P/2]}\frac{(-1)^Ne^{-il_k\cdot \mathcal{P}_N^E}}{2^s(4\pi)^{d/2}\Gamma(N)}
\int_0^\infty\rd\lambda\big\{\underbrace{\delta\cdots\delta}_{s}\mathcal{P}_N^{\prime E}\cdots \mathcal{P}_N^{\prime E}\big\}^{\mu_1\cdots\mu_P} \lambda^{N-s-\frac{d}{2}-1}e^{-\lambda \mathcal{M}_N^{E,2}-\frac{l_k^2}{4\lambda}}\ ,\notag
\end{align}
with $\mathcal{P}_N^{\prime E}\equiv \frac{il_k}{2\lambda}-\mathcal{P}_N^{ E}=\frac{inL }{2\lambda}-\mathcal{P}_N^{ E}$. The
above equation can be rewritten as
\begin{align}
\bigg\{\cdots\bigg\}_E=&
\sum_{s=0}^{[P/2]}\sum_{r=0}^{P-2s}\frac{(-1)^Ne^{-il_k\cdot \mathcal{P}_N^E}}{2^s(4\pi)^{d/2}\Gamma(N)}
\big\{\underbrace{\delta\cdots\delta}_{s}\mathcal{P}_N^{E}\cdots \mathcal{P}_N^{E}\underbrace{n\cdots n}_r\big\}^{\mu_1\cdots\mu_P}
\notag\\
&\times\left(\frac{iL}{2}\right)^r(-1)^{P-2s-r}\int_0^\infty\rd\lambda \lambda^{N-s-r-\frac{d}{2}-1}e^{-\lambda \mathcal{M}_N^{E,2}-\frac{l_k^2}{4\lambda}}\ ,
\end{align}
where the rank-$P$ tensor has now been moved out of the $\lambda$ integral. In view of Eq.~\eqref{Eq:Bessel1}, the $\lambda$ integral can be readily expressed in terms of the modified Bessel functions. Further, we change the above equation to the Minkowski space and substitute it back to Eq.~\eqref{eq:Min}. The result is 
\begin{align}\label{eq:TwithPN}
\widetilde{T}^{N,\mu_1,\cdots,\mu_P}=&\sum_{\mathbf{n}\neq 0}\sum_{s=0}^{[P/2]}\sum_{r=0}^{P-2s}
\frac{(-1)^{N+P-s-r}}{2^s(4\pi)^{d/2}\Gamma(N)}
\left(\frac{iL}{2}\right)^r \int_0^1\rd\mathcal{X}_N
\big\{\underbrace{g\cdots g}_{s}\mathcal{P}_N\cdots \mathcal{P}_N\underbrace{n\cdots n}_r\big\}^{\mu_1\cdots\mu_P}\notag\\
&\times
e^{il_k\cdot \mathcal{P}_N}\mathcal{K}_{N-s-r-\frac{d}{2}}(\frac{|\mathbf{n}|^2L^2}{4},\mathcal{M}_N^2)\ ,
\end{align}
where $\mathcal{K}_z$ is the modified Bessel function. The expression of $\mathcal{P}_N$ can be explicitly written as
\begin{align}
\mathcal{P}_N=\sum_{j=1}^{N-1}X_N^jp_j\ ,\quad 
X_N^j=\left\{
\begin{array}{ll}
  x_{N-1}\cdots x_{j+1}(1-x_j)   &  \text{for $N-1\geq j+1$} \\
   1-x_j  & \text{otherwise}
\end{array}
\right.
, \label{eq:mathP2}
\end{align}
which is obtained through Eq.~\eqref{eq:mathP}. One can rearrange Eq.~\eqref{eq:TwithPN} by inserting Eq.~\eqref{eq:mathP2} and get
\begin{align}
\widetilde{T}^{N,\mu_1,\cdots,\mu_P}=&\sum_{\mathbf{n}\neq0}
\sum^{[P/2]}_{s=0}\sum^{P-2s}_{r=0}
\sum^{N-1}_{\begin{subarray}{c}
i_{2s+1}=1\\
\cdots \\
i_{P-2s-r}=1
\end{subarray}}
\{\underbrace{g \cdots g}_{s} p \cdots p 
\underbrace{n \cdots n}_{r}
\}^{\mu_1\mu_2\cdots\mu_P}_{i_{2s+1},\cdots, i_{P-2s-r}}\frac{(-1)^{N+P-s-r}}{(4\pi)^{d/2} 2^s}
\bigg(\frac{iL}{2}\bigg)^r
\notag \\
&
\times 
\int^1_0 \rd X_N X^{i_{2s+1}}_N \cdots X^{i_{P-2s-r}}_N~e^{il_k\cdot \mathcal{P}_N}
\mathcal{K}_{N-s-r-\frac{d}{2}}(\frac{|\mathbf{n}|^2L^2}{4}, \mathcal{M}^2_N)
\ , 
\end{align}
with $\int\rd X_N\equiv \frac{1}{\Gamma(N)}\int_0^1\rd \mathcal{X}_N=\int_0^1\rd x_1\cdots\int_0^1\rd x_{N-1}x_2\cdots x_{N-1}^{N-2} $. 

Eventually, by comparing with Eq.~\eqref{eq:FVC.int.def-1} and Eq.~\eqref{eq:FVC.int.def}, we obtain a general expression for the coefficients, which reads  
\begin{align}
\vecT_{\underbrace{\scriptstyle 0\cdots 0}_{2s}i_{2s+1}\cdots i_{P-2s-r}\underbrace{\scriptstyle N\cdots N}_{r}} 
&=
\frac{2}{(4\pi)^{d/2}}
\frac{(-1)^{N+P-s-r}}{2^s}
\bigg(\frac{iL}{2}\bigg)^r
\int^1_0 \rd X_N X^{i_{2s+1}}_N \cdots X^{i_{P-2s-r}}_N~e^{iL\, n\cdot \mathcal{P}_N}
\notag \\
&
\times 
\bigg( 
\frac{|\mathbf{n}|^2L^2}{4\mathcal{M}^2_N}
\bigg)^{\frac{N-s-r-{d}/{2}}{2}}
K_{|N-s-r-\frac{d}{2}|}(|{\bf n}|L\mathcal{M}_N)
\ ,\label{eq:coe.num}
\end{align}
where $\mathcal{M}_N\equiv\sqrt{\mathcal{M}_N^2}$.
Here $K_z(Y)$ is the modified Bessel function of the second kind~\eqref{Eq:Bessel1}, which possesses the property of $K_z(Y)=K_{-z}(Y)$. It can be seen that the Lorentz invariance is broken by $n\cdot \mathcal{P}_N$ in the exponent. One can consult Appendix~\ref{app.coeff} for the tensor decompositions of $1$-, $2$-, $3$- and $4$-point tensor integrals up to rank $4$ or $5$. Explicit expressions of the tensor coefficients showing up in those decompositions are shown as well.

\section{Reduction of tensor coefficients}\label{sec:reduction}

In this section, it will be shown that, e.g., in the CM frame, the tensor decomposition of Eq.~\eqref{eq:FVC.int.def} can be further simplified into a more compact form. This is achieved by promoting the rank-$P$ tensor to a form irrelevant to the unit space-like vector. Consequently, the finite-volume sum over $\mathbf{n}$ can be done prior to the others, resulting in the $\mathbf{n}$-independent tensor coefficients. Then, we find that it is feasible to apply the procedure of PV reduction to the $\mathbf{n}$-independent tensor coefficients. General recurrence relations are obtained. 

\subsection{CM frame}
In practice, it is usually convenient to compute FVC in the rest frame of a decaying particle or in the CM frame of a scattering system. In these frames, the net three momentum is zero. Here, we consider a more restricted case that
\begin{align}\label{eq:condition}
l_k\cdot p_i=0 \Longleftrightarrow n\cdot p_i=0
\ ,\qquad i=1,\cdots,N-1\ .
\end{align}
Following Ref.~\cite{Bijnens:2013doa}, the frame with the above constraint is also referred to as CM frame. This condition can be satisfied by, e.g., elastic two-body forward scattering at threshold, mass renormalization in the rest frame, etc..

In the CM frame, the  $\widetilde{L}^{\mu_1\cdots\mu_P}$ tensors with the odd numbers of $n$-vectors vanish.
On the contrary, terms with the even numbers of $n$-vectors may survive. See Appendix~\ref{app.proof} for a detailed proof. Consequently, for the rank-$P$ tensor, its dependence on $\mathbf{n}$ can be relieved. This can be easily accomplished by employing Eq.~\eqref{eq:finalform}, i.e.
\begin{align}\label{eq:tensortoscalar}
\sum_{\mathbf{n}\neq 0}n^{\mu_1} \cdots n^{\mu_{2t}} F(n^2)=\frac{1}{2^t({d_s}/{2})_t}\left\{h\cdots h\right\}^{\mu_1\cdots\mu_{2t}}\sum_{\mathbf{n}\neq 0} (n^2)^t F(n^2)\ ,
\end{align}
where $d_s\equiv d-1$, $(d_s/2)_t$ is the Pochhammer symbol and $F(n^2)$ is an arbitrary function with respect to the module of $\mathbf{n}$. Furthermore, the auxiliary tensor $h_{\mu\nu}$ is defined,
\begin{align}
h_{\mu\nu}\equiv g_{\mu\nu}-\bar{h}_\mu \bar{h}_\nu={\rm diag}(0,-1,-1,-1)\ ,\quad \bar{h}_\mu=(1,0,0,0)\ .
\end{align}
The effect of contracting a four vector with the $h$-tensor is to eliminate the zero-th component of the vector. For more information on the properties of the $h$-tensor, we refer the readers to Appendix~\ref{app.h.tensor}. Due to the introduction of the $h$ metric tensor, the rank-$P$ tensor is now irrelevant of $\mathbf{n}$, enabling us to perform the sum over $\mathbf{n}$ in advance. Consequently, the tensor decomposition of the FVC integrals becomes
\begin{align}
\widetilde{T}^{N,\mu_1\mu_2\cdots\mu_P}
&=
\sum^{[\frac{P}{2}]}_{s=0}
\sum^{[\frac{P-2s}{2}]}_{t=0}
\sum^{N-1}_{
\begin{subarray}{c}
i_{2s+1}=1\\
\cdots \\
i_{P-2s-2t}=1
\end{subarray}
}
\big\{
\underbrace{g\cdots g}_{s}
p\cdots p 
\underbrace{h\cdots h}_{t}
\big\}^{\mu_1\cdots \mu_P}_{i_{2s+1}\cdots i_{P-2s-2t}}
\widetilde{T}^N_{\underbrace{\scriptstyle 0\cdots 0}_{2s}i_{2s+1}\cdots i_{P-2s-2t}
\underbrace{\scriptstyle N\cdots N}_{2t}}
\ ,\label{eq:CM.decom}
\end{align}
where the $\mathbf{n}$-independent coefficients are given by
\begin{align}
\widetilde{T}^N_{\underbrace{\scriptstyle 0\cdots 0}_{2s}i_{2s+1}\cdots i_{P-2s-2t}
\underbrace{\scriptstyle N\cdots N}_{2t}}
=\frac{1}{2^t(d_s/2)_t}\sum_{\mathbf{n}\neq 0}\bigg[(n^2)^t\,\vecT^{N}_{\underbrace{\scriptstyle 0\cdots 0}_{2s}i_{2s+1}\cdots i_{P-2s-2t}
\underbrace{\scriptstyle N\cdots N}_{2t}}\bigg]\ .
\label{eq:CM.num}
\end{align}
Some explicit examples are relegated to Appendix~\ref{app.coeff.cm}. Because the series in the sum rely merely on $n^2$, the triple sum can be replaced by a single sum over $n_s\equiv n_1^2+n_2^2+n_3^2$. Nonetheless, the multiplicity of $\mathbf{n}$ for a given $n_s$, denoted by $\vartheta(n_s)$, should be taken into account. Therefore, one has
\begin{align}
\widetilde{T}^N_{\underbrace{\scriptstyle 0\cdots 0}_{2s}i_{2s+1}\cdots i_{P-2s-2t}
\underbrace{\scriptstyle N\cdots N}_{2t}}
=\frac{(-1)^t}{2^t(d_s/2)_t}\sum_{n_s> 0}\bigg[\vartheta(n_s)\,n_s^t\,\vecT^{N}_{\underbrace{\scriptstyle 0\cdots 0}_{2s}i_{2s+1}\cdots i_{P-2s-2t}
\underbrace{\scriptstyle N\cdots N}_{2t}}\bigg]\ .\label{eq:Ttilde}
\end{align}
For easy reference, the values of multiplicity $\vartheta(n_s)$ for $n_s\leq 40$ are compiled in Table~\ref{tab:mult}.

\begin{table}[htbp]
\centering
\renewcommand{\tabcolsep}{0.35pc}
\begin{tabular}{c|cccccccccccccccccccc
}
\hline \hline
$n_s$         &$1$   &$2$  &$3$  &$4$ &$5$ &$6$ &$7$ &$8$ &$9$ &$10$ &$11$ &$12$ &$13$ &$14$ &$15$ &$16$ &$17$ &$18$ &$19$ &$20$ 
\\
$\vartheta(n_s)$  &$6$  &$12$ &$8$ &$6$ &$24$ &$24$ &$0$ &$12$ &$30$ &$24$ &$24$ &$8$ &$24$ &$48$ &$0$ &$6$ &$48$ &$36$ &$24$ &$24$
\\
\hline 
$n_s$  &$21$ &$22$ &$23$ &$24$ &$25$ &$26$ &$27$ &$28$ &$29$ &$30$ &$31$ &$32$ &$33$ &$34$ &$35$ &$36$ &$37$ &$38$ &$39$ &$40$  \\
$\vartheta(n_s)$  &$48$ &$24$ &$0$ &$24$ &$30$ &$72$ &$32$ &$0$ &$72$ &$48$ &$0$ &$12$ &$48$ &$48$ &$48$ &$30$ &$24$ &$72$ &$0$ &$24$
\\
\hline \hline
\end{tabular}
\caption{The values of multiplicity up to $n_s=40$.}
\label{tab:mult}
\end{table}

It is worth noting that, in some regularization schemes other than dimensional regularization, momentum integration is performed with a regulator in order to avoid UV divergences in the infinite volume. Correspondingly, at finite volume there usually exists an upper limit for $n_s$, see e.g.~\cite{Doring:2011vk} for more discussions. Specifically, if a cut-off $\Lambda$ is adopted to restrict the momentum in the integration, the upper limit $n_{\rm max}$ of the summation in Eq.~\eqref{eq:Ttilde} is given by
$
n_s^{\rm max}={\rm Ceiling}[\frac{\Lambda^2L^2}{4\pi^2}]\ . 
$ 

More generally, for large argument the Bessel functions in the sum of
 Eq.~\eqref{eq:Ttilde} decay exponentially as
\begin{align}
    K_{|N-s-2t-\frac{d}{2}|}(x)\sim \sqrt{\frac{\pi}{2x}}e^{-x}
    \ , \quad x=\sqrt{n_s}L\mathcal{M}_N\ .
\end{align}
Given that $L\mathcal{M}_N\gg 1$, the sum converges rapidly. Therefore, the sum can be truncated at a certain value of $n_s$, provided that a good approximation is generated for the complete sum~\cite{Colangelo:2003hf}. In another word, there is no need to sum $n_s$ to infinity in most cases.

\subsection{Passarino-Veltman reduction}
As discussed above, in the CM frame only even numbers of $n$-vectors are left, of which every two are replaced by an auxiliary tensor $h_{\mu\nu}$. Under this circumstance, the PV reduction is still valid~\cite{Passarino:1978jh}, although the Lorentz invariance is absent. The essence of PV reduction is to establish algebraic relations between the tensor coefficients, by means of contracting the tensor integrals, i.e. Eq.~\eqref{eq:tensor.int.def0}, with external momenta $p_{i\mu}$ and the metric tensor $g_{\mu\nu}$. For more details, the readers are referred to Refs.~\cite{Passarino:1978jh,Denner:2002ii,Denner:2005nn}. The relations at finite volume are more complicated.

Let us start with the simplest situation. That is to deal with the one-point tensor integrals. It is obvious, as can be seen from Eq.~\eqref{eq:tensor.int.def0}, that they are irrelevant to external momenta. Therefore, they can only be contracted by the metric tensor, leading to reduction of tensor rank and cancellation of denominators. After the decomposition equations~\eqref{eq:CM.decom} are imposed, the recurrence relations can be read off by comparing coefficients, 
\begin{align}
\big[(d-1)+2(t-1)\big]
\widetilde{A}_{\underbrace{\scriptstyle0\cdots0}_{2s}\underbrace{\scriptstyle 1\cdots 1}_{2t}}
+
\big[ 
d+2s
+4(t-1)
\big]
\widetilde{A}_{\underbrace{\scriptstyle0\cdots0}_{2s+2}\underbrace{\scriptstyle 1\cdots 1}_{2t-2}}
=
m^2_1 \widetilde{A}_{\underbrace{\scriptstyle0\cdots0}_{2s}\underbrace{\scriptstyle 1\cdots 1}_{2t-2}}
\ .\label{eq:1-point.for}
\end{align}  
Specifically, the relations up to rank $6$ are given below.
\begin{align}
d\widetilde{A}_{00}+(d-1)\widetilde{A}_{11}&=m^2_1 \widetilde{A}_{0}
\ , \\
(d+2)\widetilde{A}_{0000}+(d-1)\widetilde{A}_{0011}&=m^2_1\widetilde{A}_{00}
\ ,\\
(d+4)\widetilde{A}_{0011}+(d+1)\widetilde{A}_{1111}&=m^2_1\widetilde{A}_{11}
\ ,
\\
(d+4)\widetilde{A}_{000000}+(d-1)\widetilde{A}_{000011}&=m^2_1\widetilde{A}_{0000}
\ , \\
(d+6)\widetilde{A}_{000011}+(d+1)\widetilde{A}_{001111}&=m^2_1\widetilde{A}_{0011}
\ , \\
(d+8)\widetilde{A}_{001111}+(d+3)\widetilde{A}_{111111}&=m^2_1\widetilde{A}_{1111}
\ .
\end{align}
All the above relations can either be checked numerically by inserting Eq.~\eqref{eq:coe.num} and Eq.~\eqref{eq:CM.num}, or be verified by the recurrence relations of the modified Bessel functions $K_z(Y)$, Eq.~\eqref{eq:K.rec.rel}.

\begin{figure}[tbhp]
    \centering
    \includegraphics[scale=0.4]{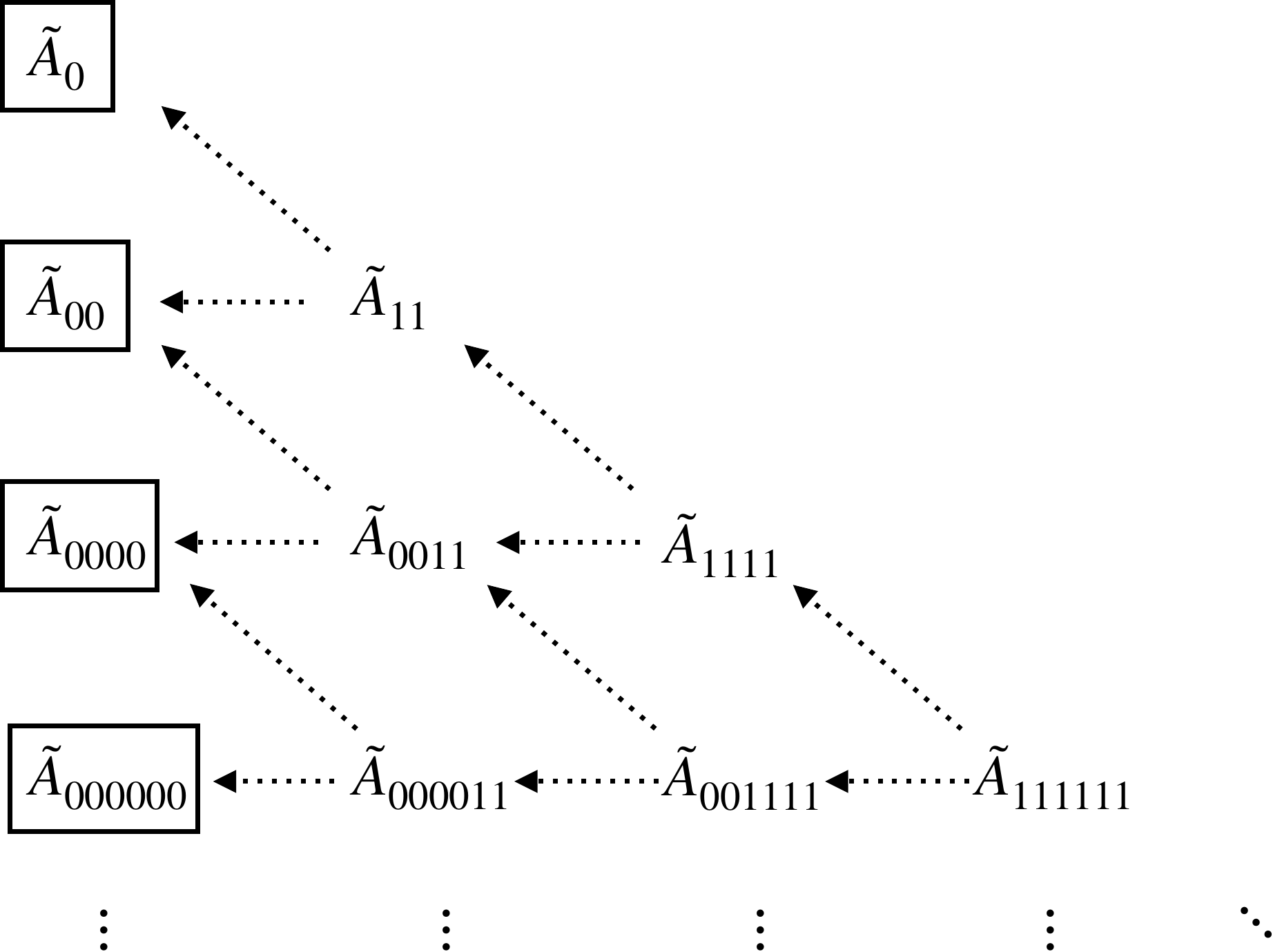}
    \caption{Schematic roadmap for PV reduction of one-point FVC tensor integrals.  The dashed lines with arrows represent simplification operations by the recursive use of Eq.~\eqref{eq:1-point.for}.  The coefficients on the left, enclosed in the boxes, are regarded as tensor basis. }
    \label{fig:A0s}
\end{figure}

In the infinite volume, one-point tensor integrals can be uniformly expressed in terms of the scalar integral $A_0$, see e.g. Eq.~(3.4) of Ref.~\cite{Denner:2005nn}. It is also interesting to discuss the analogue for the case of FVC. To that end, it can be found from Fig.~\ref{fig:A0s} that all the one-loop FVC integrals can be reduced to a linear combination of $\widetilde{A}_{\underbrace{\scriptstyle 0\cdots0}_{2s}}$ with $s=0,1,2,\cdots$. Thus, the $\widetilde{A}_{0}$, $\widetilde{A}_{00}$, $\widetilde{A}_{0000}$, etc., can be adopted as the tensor basis. More specifically, the tensor coefficients of one-point integrals can be expressed by
\begin{align}
\widetilde{A}_{\underbrace{\scriptstyle 0\cdots0}_{2s}\underbrace{\scriptstyle 1\cdots1}_{2t}}
=\sum_{i=0}^t\bigg\{\frac{[m_1^2]^{t-i}}{\prod_{j=1}^ta(j)}
\sum_{
\begin{subarray}{c}
i_{1}=0\\
\cdots \\
i_{t}=0
\end{subarray}
}^1\bigg[\delta_{i,\sum_{j=1}^ti_j}\prod_{j=1}^t[b(j)]^{i_j}\bigg]\widetilde{A}_{\underbrace{\scriptstyle 0\cdots0}_{2(s+i)}}\bigg\}\ ,
\end{align}
where $a(j)=(d-1)+2(j-1)$, $b(j)=-[d+2s+4(j-1)]$, and $\delta$ is the Kronecker delta.

\begin{figure}[tbhp]
    \centering
    \includegraphics[scale=0.4]{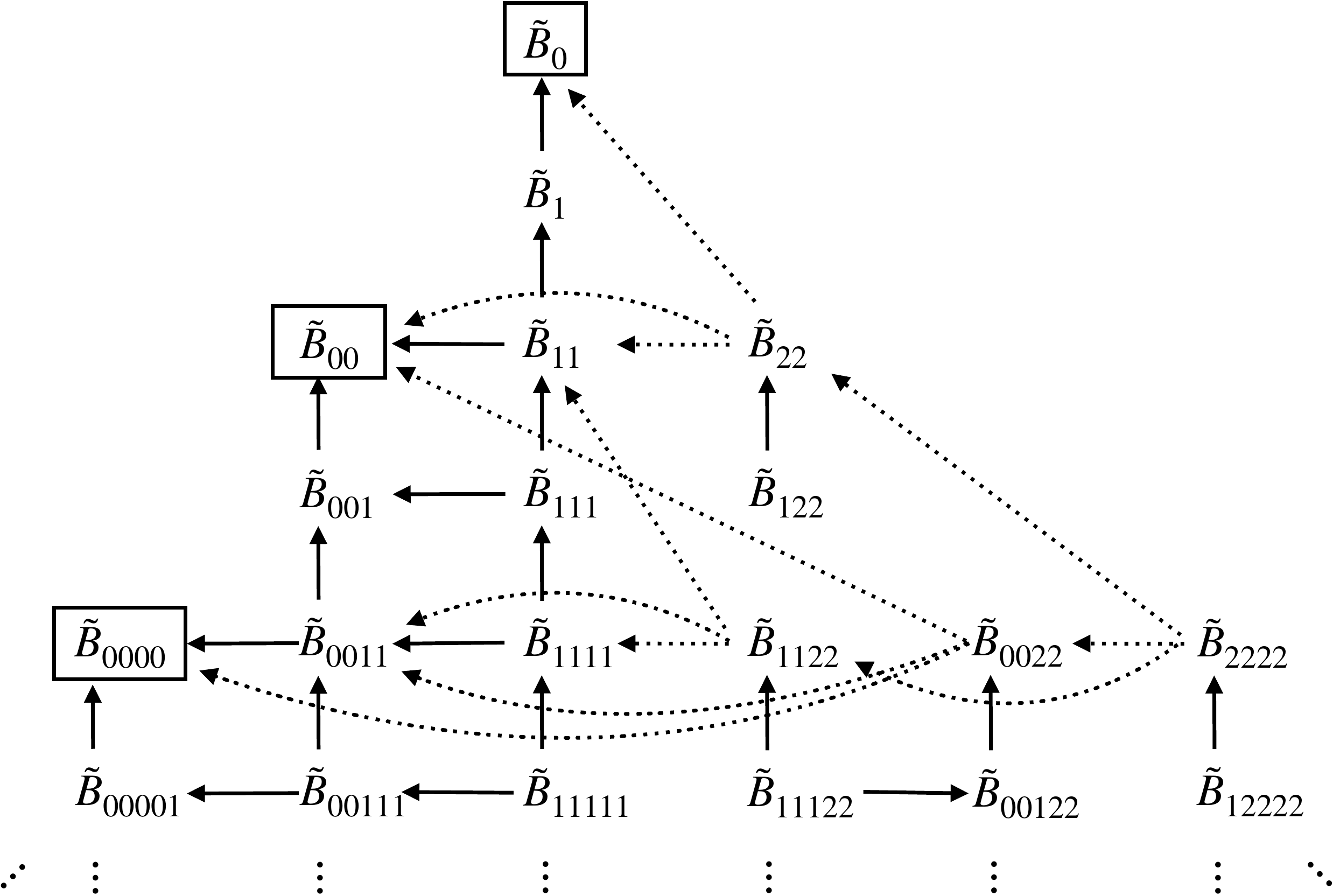}
    \caption{Schematic roadmap for PV reduction of two-point FVC tensor coefficients. The solid and dashed lines stand for the recursive applications of Eq.~\eqref{eq:B.PVeqs2} and Eq.~\eqref{eq:B.PVeqs1}, respectively. The boxed tensor coefficients are taken as the basis. }
    \label{fig:B0s}
\end{figure}

As for two-point integrals, the contraction of the external momentum $p_{1\mu}$ makes us obtain the equation as
\begin{align}
&2p^2_1\widetilde{B}_{\underbrace{\scriptstyle0\cdots0}_{2s}\underbrace{\scriptstyle1\cdots1}_{\ell+1}\underbrace{\scriptstyle 2\cdots2}_{2t}}
+
2\ell\widetilde{B}_{\underbrace{\scriptstyle0\cdots0}_{2s+2}\underbrace{\scriptstyle1\cdots1}_{\ell-1}\underbrace{\scriptstyle 2\cdots2}_{2t}}\nonumber\\
=&
\delta_{\ell0}\widetilde{A}_{\underbrace{\scriptstyle0\cdots0}_{2s}\underbrace{\scriptstyle 1\cdots1}_{2t}}(2)
-(-1)^{\ell}\widetilde{A}_{\underbrace{\scriptstyle0\cdots0}_{2s}\underbrace{\scriptstyle 1\cdots1}_{2t}}(1)
-
f_2\widetilde{B}_{\underbrace{\scriptstyle0\cdots0}_{2s}\underbrace{\scriptstyle1\cdots1}_{\ell}\underbrace{\scriptstyle 2\cdots2}_{2t}}
\ ,\label{eq:B.PVeqs2}
\end{align}
with $f_2=p_1^2-m_2^2+m_1^2$. Here $\widetilde{A}(k)$ means that it is obtained by omitting the $k$-th denominator in Eq.~\eqref{eq:tensor.int.def0}. On the other hand, an extra relation can be deduced by contracting the metric tensor, which reads
\begin{align}
&[d+2s+2\ell+4t]\widetilde{B}_{\underbrace{\scriptstyle0\cdots0}_{2s+2}\underbrace{\scriptstyle1\cdots1}_{\ell}\underbrace{\scriptstyle 2\cdots2}_{2t}}
+
[(d-1)+2t]\widetilde{B}_{\underbrace{\scriptstyle0\cdots0}_{2s}\underbrace{\scriptstyle1\cdots1}_{\ell}\underbrace{\scriptstyle 2\cdots2}_{2t+2}}
+
p^2_1
\widetilde{B}_{\underbrace{\scriptstyle0\cdots0}_{2s}\underbrace{\scriptstyle1\cdots1}_{\ell+2}\underbrace{\scriptstyle 2\cdots2}_{2t}}
\notag \\
=&
(-1)^{P-2-2(t+s)}
\widetilde{A}_{\underbrace{\scriptstyle0\cdots0}_{2s}\underbrace{\scriptstyle 1\cdots1}_{2t}}(1)
+
m^2_1 \widetilde{B}_{\underbrace{\scriptstyle0\cdots0}_{2s}\underbrace{\scriptstyle1\cdots1}_{\ell}\underbrace{\scriptstyle 2\cdots2}_{2t}}
\ . \label{eq:B.PVeqs1}
\end{align}
Explicit relations for arbitrary rank are obtainable by just setting $s$, $\ell$ and $t$ to the values as needed. For the sake of easy reference, here we show the relations for the two-point integrals up to rank $4$. 
\begin{align}
p^2_1\widetilde{B}_1+\frac{1}{2}(p^2_1+m^2_1-m^2_2)\widetilde{B}_0&=\frac{1}{2}\widetilde{A}_{0}(m^2_1)-\frac{1}{2}\widetilde{A}_{0}(m^2_2)
\ . \\
%
p^2_1\widetilde{B}_{11}+d\widetilde{B}_{00}+(d-1)\widetilde{B}_{22}&=\widetilde{A}_{0}(m^2_2)+m^2_1\widetilde{B}_{0}
\ , \\
p^2_1\widetilde{B}_{11}+\widetilde{B}_{00}+\frac{1}{2}(p^2_1+m^2_1-m^2_2)\widetilde{B}_{1}&=
\frac{1}{2}\widetilde{A}_{0}(m^2_2)
\ . \\
%
p^2_1\widetilde{B}_{111}+(d+2)\widetilde{B}_{001}+(d-1)\widetilde{B}_{122}&=m^2_1\widetilde{B}_{1}-\widetilde{A}_{0}(m^2_2)
\ , \\
p^2_1\widetilde{B}_{111}+2\widetilde{B}_{001}+\frac{1}{2}(p^2_1+m^2_1-m^2_2)\widetilde{B}_{11}&=-\frac{1}{2}\widetilde{A}_{0}(m^2_2)
\ , \\
p^2_1\widetilde{B}_{001}+\frac{1}{2}(p^2_1+m^2_1-m^2_2)\widetilde{B}_{00}&=\frac{1}{2}\big[\widetilde{A}_{00}(m^2_1)-\widetilde{A}_{00}(m^2_2)\big]
\ , \\
p^2_1\widetilde{B}_{122}+\frac{1}{2}(p^2_1+m^2_1-m^2_2)\widetilde{B}_{22}&=\frac{1}{2}\big[\widetilde{A}_{11}(m^2_1)-\widetilde{A}_{11}(m^2_2)\big]
\ . \\
%
p^2_1\widetilde{B}_{1111}+(d+4)\widetilde{B}_{0011}+(d-1)\widetilde{B}_{1122}&=\widetilde{A}_{0}(m^2_2)+m^2_1\widetilde{B}_{11}
\ , \\
p^2_1\widetilde{B}_{0011}+(d+2)\widetilde{B}_{0000}+(d-1)\widetilde{B}_{0022}&=\widetilde{A}_{00}(m^2_2)+m^2_1\widetilde{B}_{00}
\ , \\
p^2_1\widetilde{B}_{1122}+(d+4)\widetilde{B}_{0022}+(d+1)\widetilde{B}_{2222}&=\widetilde{A}_{11}(m^2_2)+m^2_1\widetilde{B}_{22}
\ , \\
p^2_1\widetilde{B}_{1111}+3\widetilde{B}_{0011}+\frac{1}{2}(p^2_1+m^2_1-m^2_2)\widetilde{B}_{111}&=\frac{1}{2}\widetilde{A}_{0}(m^2_2) 
\ , \\
p^2_1\widetilde{B}_{0011}+\widetilde{B}_{0000}+\frac{1}{2}(p^2_1+m^2_1-m^2_2)\widetilde{B}_{001}&=\frac{1}{2}\widetilde{A}_{00}(m^2_2)
\ , \\
p^2_1\widetilde{B}_{1122}+\widetilde{B}_{0022}+\frac{1}{2}(p^2_1+m^2_1-m^2_2)\widetilde{B}_{122}&=\frac{1}{2}\widetilde{A}_{11}(m^2_2)
\ .
\end{align}
It should be noted that some of the above relations have been given in Ref.~\cite{Bijnens:2014dea}, though different notations are adopted therein. It can be found that, by solving the above equations, the tensor coefficients of two-point integrals can be reduced to the ones with lower rank and one-point tensor coefficients. 
A typical reduction procedure is graphically illustrated in Fig.~\ref{fig:B0s}. The number of subscripts \lq\lq$2$" is reduced by recursively utilizing Eq.~\eqref{eq:B.PVeqs1}, which is represented by the dashed lines in Fig.~\ref{fig:B0s}. Meanwhile, the indices \lq\lq$1$" can be eliminated by making use of Eq.~\eqref{eq:B.PVeqs2}, and the corresponding procedure is indicated by the solid lines in Fig.~\ref{fig:B0s}. In the end, only those tensor coefficients with even numbers of \lq\lq$0$" survive, just like the case for one-point integrals.

Likewise, the reduction of the coefficients of $N$-point ($N>3$) tensor integrals can be achieved recursively as well. The reduction procedure of two-point tensor coefficients can be straightforwardly extended to the $N$-point cases.  

For a given $N$-point rank-$P$ tensor integral, the contraction of momentum $p_j^{\mu_1}$ yields the following relationship of the coefficients 
\begin{align}
&2\sum_{r=2s+1}^{P-1-2s-2t}\delta_{ji_r}\,\widetilde{T}^N_{\underbrace{\scriptstyle 0\cdots 0}_{2(s+1)}i_{2s+1}\cdots\slashed{i}_r\cdots i_{P-1-2s-2t}\underbrace{\scriptstyle N\cdots N}_{2t}}
+
\sum_{m=1}^{N-1}Z_{jm}^{(N-1)}\,\widetilde{T}^N_{\underbrace{\scriptstyle 0\cdots0}_{2s}mi_{2s+1}\cdots i_{P-1-2s-2t}\underbrace{\scriptstyle N\cdots N}_{2t}}\notag\\
=&\,\bar{\delta}_{ji_{2s+1}}\cdots
\bar{\delta}_{ji_{P-1-2s-2t}}\,\widetilde{T}^{N-1}_{\underbrace{\scriptstyle 0\cdots 0}_{2s}(i_{2s+1})_j\cdots (i_{P-1-2s-2t})_j\underbrace{\scriptstyle N-1\cdots N-1}_{2t}}(j+1)\notag\\
&-
\widetilde{T}^{\prime N-1}_{\underbrace{\scriptstyle 0\cdots 0}_{2s}
i_{2s+1}\cdots i_{P-1-2s-2t}\underbrace{\scriptstyle N\cdots N}_{2t}
}(1)-f_{j+1}\widetilde{T}^{ N}_{\underbrace{\scriptstyle 0\cdots 0}_{2s}
i_{2s+1}\,\cdots i_{P-1-2s-2t\underbrace{\scriptstyle N\cdots N}_{2t}}
}\ ,\label{eq.PV.npts1}
\end{align}
where $\bar{\delta}_{ij}\equiv 1-\delta_{ij}$, $f_j\equiv p_{j-1}^2-m_j^2+m_1^2$ and $i_r\in \{1,\cdots, N-1\}$. In the first term of the above equation, a slashed index is to indicate that the index is omitted. Here, $(i_r)_j=i_r$ if $j>i_r$, and $(i_r)_j=i_r-1$ if $j<i_r$. It is worth noting that the above relation for FVC is in the same form as the one in the infinite volume, given in e.g. Ref.~\cite{Denner:2005nn}. This is due to the fact that no extra terms are generated in our case of FVC, thanks to the identity $p_j^{\mu_1}h_{\mu_1\mu_2}=0$. The Gram matrix is defined by
\begin{align}
Z^{(N-1)}=
\left(
\begin{array}{ccc}
   2p_1\cdot p_1  & \cdots & 2p_1\cdot p_{N-1} \\
    \vdots &  \ddots & \vdots\\
2p_{N-1}\cdot p_1 &  \cdots & 2p_{N-1}\cdot p_{N-1}
\end{array}
\right)\ . 
\label{eq:GramM}
\end{align}

Besides, the contraction of $g^{\mu_1\mu_2}$ leads to the equation
\begin{align}
&[d+2s+2\ell+4t]\widetilde{T}^{N}_{\underbrace{\scriptstyle 0\cdots0}_{2(s+1)}i_{2s+1}\cdots i_{P-2-2s-2t}\underbrace{\scriptstyle N\cdots N}_{2t}}
+[(d-1)+2t]\widetilde{T}^{N}_{\underbrace{\scriptstyle 0\cdots0}_{2s}i_{2s+1}\cdots i_{P-2-2s-2t}\underbrace{\scriptstyle N\cdots N}_{2(t+1)}}
\notag\\
&
+\frac{1}{2}\sum_{n,m=1}^{N-1}Z^{(N-1)}_{nm}\widetilde{T}^{N}_{\underbrace{\scriptstyle 0\cdots0}_{2s}nmi_{2s+1}\cdots i_{P-2-2s-2t}\underbrace{\scriptstyle N\cdots N}_{2t}}\notag\\
=&\,\widetilde{T}^{\prime,\,N-1}_{\underbrace{\scriptstyle 0\cdots0}_{2s}\scriptstyle i_{2s+1}\cdots i_{P-2-2s-2t}\underbrace{\scriptstyle N\cdots N}_{2t}}(1)
+m_1^2\,\widetilde{T}^{N}_{\underbrace{\scriptstyle 0\cdots0}_{2s}i_{2s+1}\cdots i_{P-2-2s-2t}\underbrace{\scriptstyle N\cdots N}_{2t}}\ ,\label{eq.PV.npts2}
\end{align}
with $\ell=P-2-2s-2t$.

\begin{figure}[tbhp]
\centering
\includegraphics[scale=0.45]{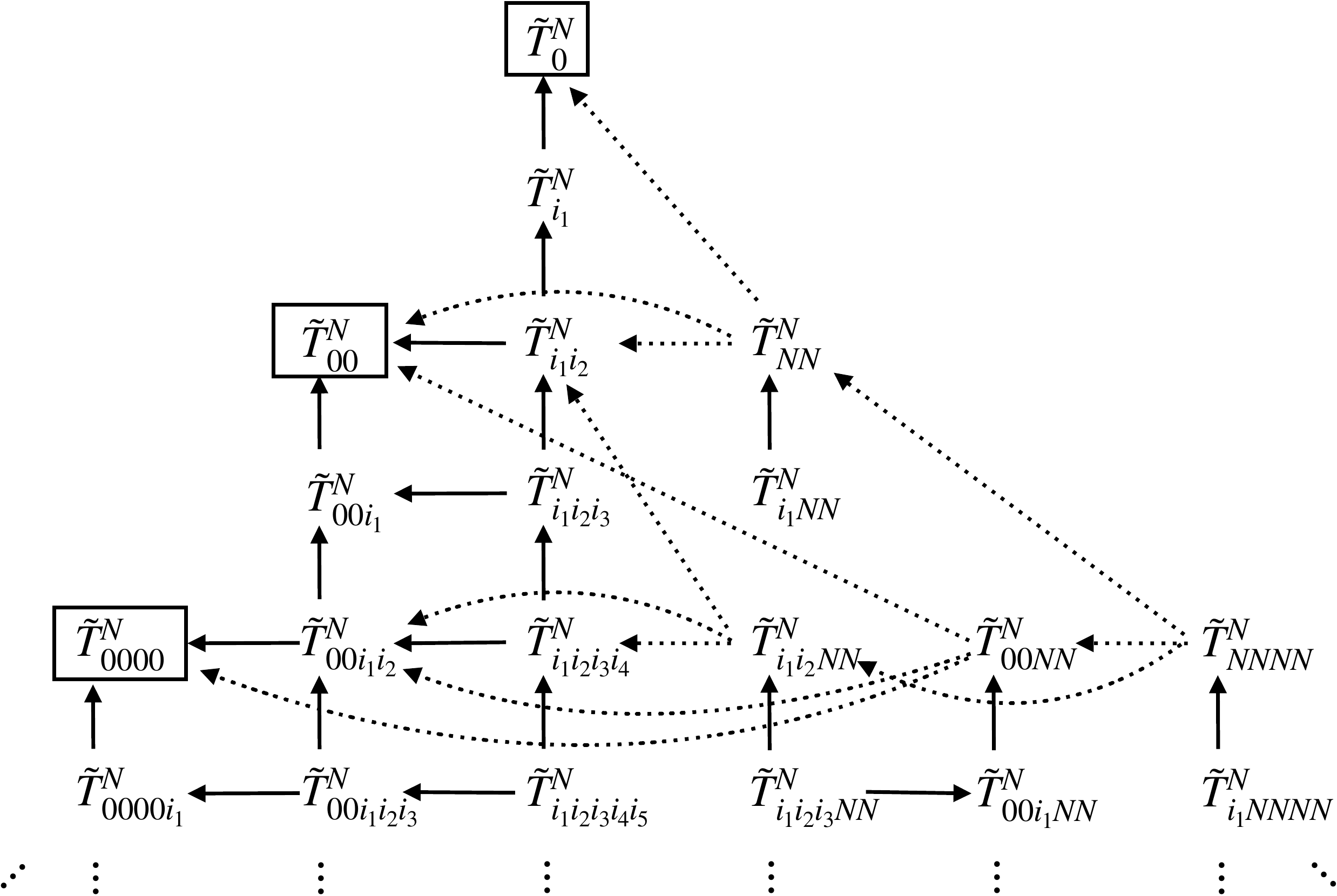}
\caption{Schematic roadmap for PV reduction of $N$-point FVC tensor coefficients. The solid and dashed lines stand for the recursive applications of Eq.~\eqref{eq.PV.npts1} and Eq.~\eqref{eq.PV.npts2}, respectively. The boxed tensor coefficients are taken as the basis.}
\label{fig:N0s}
\end{figure}

In Eq.~\eqref{eq.PV.npts1} and Eq.~\eqref{eq.PV.npts2}, the coefficients of the non-standard tensor integral, labelled by a prime, can be re-expressed in terms of the ones of the standard tensor integral, i.e.,
\begin{align}
&\widetilde{T}^{\prime N-1}_{\underbrace{\scriptstyle0\cdots 0}_{2s}\underbrace{\scriptstyle 1\cdots 1}_{n}i_{2s+n+1}\cdots i_{P-2s-2t-n}\underbrace{\scriptstyle N\cdots N}_{2t}}(1)\notag\\
=&(-1)^n\sum_{m=0}^n C_n^m\sum^{N-2}_{\scriptstyle\begin{subarray}{c}
i_1=1,\\
\cdots \\
i_m=1
\end{subarray}}
\widetilde{T}^{N-1}_{\underbrace{\scriptstyle0\cdots 0}_{2s}i_1\cdots i_mi_{2s+n+1}-1\cdots {i}_{P-2s-2t-n}-1\underbrace{\scriptstyle N-1\cdots N-1}_{2t}}(1)\ .
\end{align}
The non-standard tensor integral is defined by omitting the first denominator factor $D_1$ in Eq.~\eqref{eq:tensor.int.def0}.

Schematic illustration of PV reduction of $N$-point FVC tensor coefficients is displayed in Fig.~\ref{fig:N0s}. The solid and dashed lines denote the utilizing of Eq.~\eqref{eq.PV.npts1} and Eq.~\eqref{eq.PV.npts2}, respectively. 

It should be pointed out that we only aim at finding out the feasibility of PV reduction and the existence of a tensor basis for the one-loop integrals at finite volume. It is a first attempt and has its own limitations. For instance, the Gram matrix Eq.~\eqref{eq:GramM} may appear in the denominator during the reduction procedure. Numerical instabilities could be caused when it is small. In some special cases, the Gram matrix can even be identical to zero, and alternative reduction approaches should be employed. See e.g. Ref.~\cite{Denner:2005nn} and the references therein. Anyway, one can always use Eq.~\eqref{eq:CM.num} or Eq.~\eqref{eq:Ttilde} for a practical calculation when the above-mentioned drawbacks are encountered.

\section{A pedagogic example of application}\label{sec:application}
\begin{figure}[hptb]
    \centering
    \includegraphics[scale=0.85]{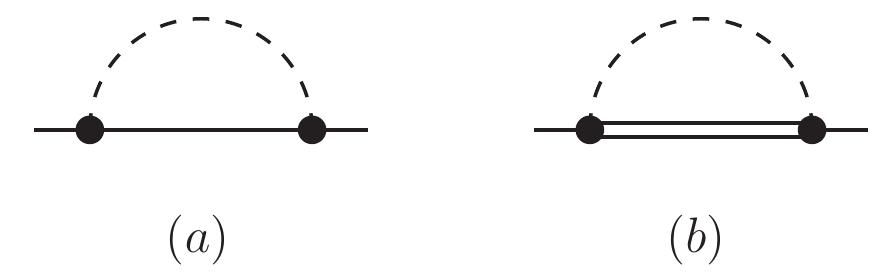}
    \caption{Leading one-loop Feynman diagrams contributing the nucleon mass in baryon ChPT with pions, nucleons and deltas as explicit degrees of freedom. The solid, dashed and double lines represent nucleons, pions and deltas, in order. The black dots denote vertices of $\mathcal{O}(p^1)$. }
    \label{fig:m}
\end{figure}

The calculation of the nucleon mass up to next-to-leading order in the chiral perturbation theory can be found elsewhere~\cite{QCDSF-UKQCD:2003hmh,Chen:2012nx,Alvarez-Ruso:2013fza,Yao:2016vbz}. For the sake of easy comparison, the explicit expressions of the one-loop self-energies, given by Ref.~\cite{Alvarez-Ruso:2013fza}, are compiled in Appendix~\ref{app.se}. Here, we are going to recalculate the FVC to the nucleon mass using the formulation of the one-loop integrals proposed in this work. The effects of FVC stem only from loop diagrams. Relevant Feynman diagrams contributing to the nucleon mass at leading one-loop order are shown in Fig.~\ref{fig:m}. The chiral effective Lagrangians can be found, e.g., in Ref.~\cite{Yao:2016vbz}, and the corresponding Feynman rules are given in Ref.~\cite{Yao:2019jyg}. Our explicit expressions of the one-loop self-energies are relegated to Appendix~\ref{app.se}.

It can be seen from Eq.~\eqref{eq.se.a.long} and Eq.~\eqref{eq.se.b.long} that it is convenient to decompose the self-energy of the nucleon into the form as
\begin{align}
\Sigma(\slashed{p},\slashed{n}) =
 \mathcal{A}+\slashed{p}\mathcal{B}+\sum_{{\bf n}\neq 0}\slashed{n}\mathcal{C} 
\ ,
\end{align}
where $\mathcal{A}$, $\mathcal{B}$ and $\mathcal{C}$ are functions with respect to scalar products of the external momentum $p^\mu$ and the unit space-like vector $n^\mu$. The occurrence of the third term is due to the introduction of spatial boundary conditions of the finite volume.  

For diagram (a), in view of Eq.~\eqref{eq.se.a.long}, one obtains 
\begin{align}
\mathcal{A}_a
&=\frac{3g_A^2m_{N}}{4F^2_{\pi}}
\sum_{\mathbf{n}\neq 0}
\bigg\{
s \vecB_0+2s \vecB_1+d \vecB_{00}+s \vecB_{11}+n^2\vecB_{22}
-2 n\cdot p \left[\vecB_2+\vecB_{12}\right]
\bigg\}\ ,\notag\\
\mathcal{B}_a
&=\frac{3g_A^2}{4F^2_{\pi}}
\sum_{\mathbf{n}\neq 0}
\bigg\{
s \vecB_1+2s \vecB_{11}+2d \vecB_{00}+
(d+2) \vecB_{001}
+s \vecB_{111}
\notag \\
& \hspace{2cm}
+n^2\left[2\vecB_{22}+\vecB_{122}\right]
-2n\cdot p\left[\vecB_2+2\vecB_{12}+\vecB_{112}\right]
\bigg\}\ ,\notag\\
\mathcal{C}_a
&=\frac{3g_A^2}{4F^2_{\pi}}
\bigg\{
s \vecB_2
-(d+2)\vecB_{002}
-s\vecB_{112}
-n^2\vecB_{222}
+2n\cdot p \vecB_{122}
\bigg\}\ ,
\end{align}
with $s\equiv p^2$. The explicit expressions of the tensor coefficients $\vecB$ can be found in Appendix~\ref{app.coeff}. Here $g_A$ is the axial coupling constant, $F_\pi$ is the pion decay constant, and $m_N$ denotes the nucleon mass in the chiral limit. The arguments of the loop functions are omitted for brevity. In the CM frame, i.e. the rest frame of the nucleon, one has $\bar{u}(p)\slashed{n}u(p)=0$, hence the term $\mathcal{C}$ is switched off. In the meantime, $\mathcal{A}$ and $\mathcal{B}$ can be simplified to
\begin{align}
\mathcal{A}_a&=\frac{3g_A^2m_{N}}{4F^2_{\pi}}\bigg\{
s \tdB_0+2s \tdB_1+d \tdB_{00}+s \tdB_{11}+(d-1)\tdB_{22}
\bigg\}\ ,\notag\\
\mathcal{B}_a&=\frac{3g_A^2}{4F^2_{\pi}}\bigg\{
s \tdB_1+2s \tdB_{11}+2d \tdB_{00}+(d+2) \tdB_{001}
+s \tdB_{111}
+(d-1)\left[2\tdB_{22}
+\tdB_{122}\right]
\bigg\}\ .
\end{align}
It can be seen that the tensor coefficients with odd numbers of indices \lq\lq2" disappear. They can be further simplified by making use of PV reduction,
\begin{align}
\mathcal{A}_a(L)&=\frac{3g_A^2m_{N}}{4F^2_{\pi}}\bigg\{
\widetilde{A}_0(m_{N}^2;L)+M_\pi^2 \tdB_0(m_{N}^2,m_{N}^2,M_\pi^2;L)
\bigg\}\ ,\notag\\
\mathcal{B}_a(L)&=\frac{1}{m_{N}}\mathcal{A}_a(L)\ ,
\end{align}
where the arguments of the loop functions are now explicitly shown. Here $M_\pi$ is the pion mass and $L$ is the size of the spatial cubic box. 

The same procedure can be conducted on diagram (b), which yields
\begin{align}
\mathcal{A}_b(L)&=-\frac{h_A^2}{3F^2_{\pi}m_\Delta}
\bigg\{
(m_\Delta^2-m_N^2+3M_\pi^2)\widetilde{A}_0(M_\pi^2;L)-(m_\Delta^2+m_N^2-M_\pi^2)\widetilde{A}_0(m_\Delta^2;L)\notag\\
&\hspace{2.5cm}+\lambda(m_\Delta^2,m_N^2,M_\pi^2) \widetilde{B}_0(m_N^2,m_\Delta^2,M_\pi^2;L)
\bigg\} \ ,
\notag\\
\mathcal{B}_b(L)&=\frac{h_A^2}{6F^2_{\pi}m_\Delta^2m_N^2}
\bigg\{
\lambda(m_\Delta^2,m_N^2,M_\pi^2)\widetilde{A}_0(m_\Delta^2;L)-[(m_\Delta^2-M_\pi^2)^2-m_N^4+4m_N^2M_\pi^2]\widetilde{A}_0(M_\pi^2;L)\notag\\
&\hspace{2.5cm}+4m_N^2[\widetilde{A}_{00}(m_\Delta^2;L)-\widetilde{A}_{00}(M_\pi^2;L)]\notag\\
&\hspace{2.5cm}+\lambda(m_\Delta^2,m_N^2,M_\pi^2)(m_\Delta^2+m_N^2-M_\pi^2)\widetilde{B}_0(m_N^2,m_\Delta^2,M_\pi^2;L)
\bigg\}\ ,
\end{align}
with $\lambda(a,b,c)=a^2+b^2+c^2-2ab-2ac-2bc$ being the K\"all\'{e}n function. The coupling constant of the $\pi N\Delta$ interaction is denoted by $h_A$, and $m_\Delta$ is mass of the delta resonance in the chiral limit. As expected, only scalar integrals and tensor coefficients, with even numbers of indices \lq\lq$0$", show up in the PV reduced results.

Eventually, the expression of the FVC on the nucleon mass reads
\begin{align}
m_N^{\rm FVC}(L)=\big[\mathcal{A}(L)+m_N \mathcal{B}(L)\big]\ ,
\end{align}
with
\begin{align}
\mathcal{A}(L)=\mathcal{A}_a(L)+\mathcal{A}_b(L)\ ,\quad \mathcal{B}(L)=\mathcal{B}_a(L)+\mathcal{B}_b(L)\ .
\end{align}
The FVC result of the nucleon mass is unique in form after PV reduction, and hence it is portable and convenient to be used elsewhere.

\begin{figure}
\centering    
\includegraphics[width=0.95\textwidth]{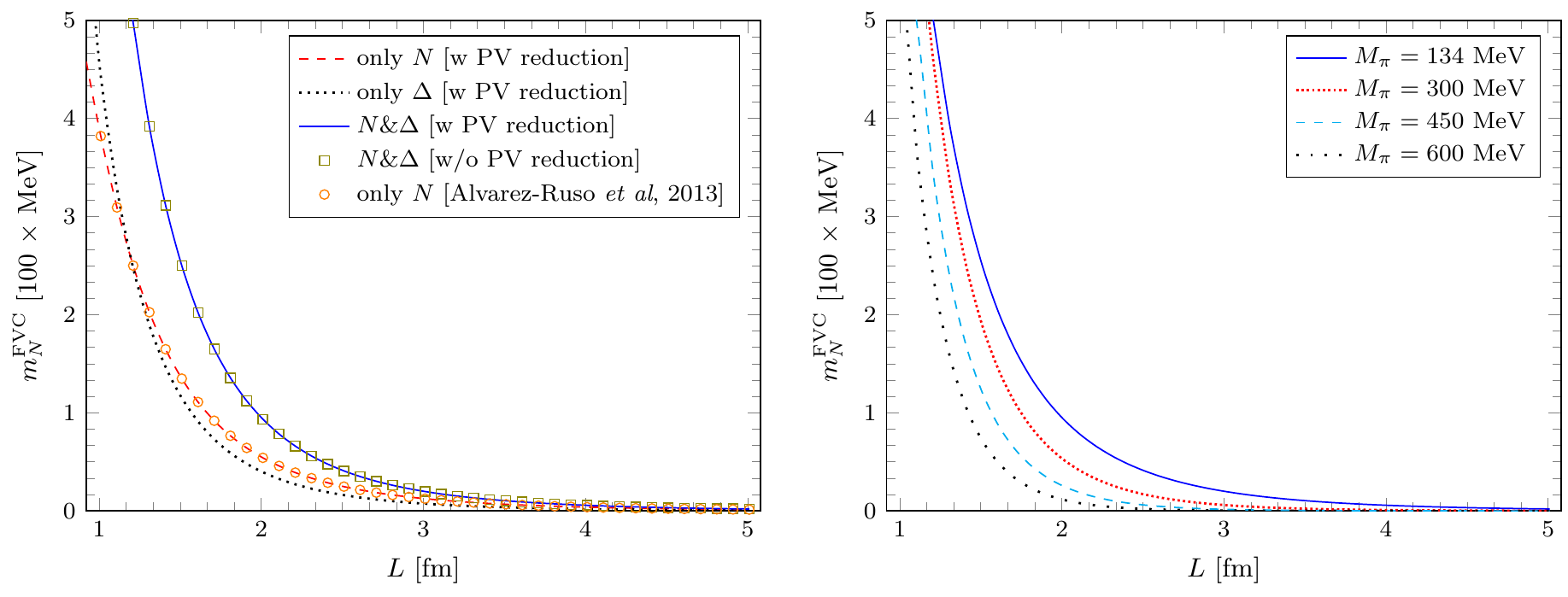}
\caption{Left panel: FVC to the nucleon mass. Right panel: the $L$-dependence of the nucleon mass with various values of the pion mass.}
\label{fig:FVC.mass}
\end{figure}

In our numerical computation, the values of the involved parameters are set to be: $g_A=1.27$, $h_A=1.35$, $F_\pi=92.2$~MeV and $M_\pi=134$~MeV. For the masses in the chiral limit, we take $m_N=890$~MeV and $m_\Delta=1170$~MeV from Ref.~\cite{Alvarez-Ruso:2013fza}. 
Our results are plotted in Fig.~\ref{fig:FVC.mass}. The red dashed, black dotted and blue solid lines represent the contributions from diagrams (a), (b) and their sum, in order. The olive open squares stand for the total contribution that is calculated by using the expressions without PV reduction. The validity of the PV reduction for the FVC tensor coefficients is explicitly verified, due to the fact that the numerical results with and without PV reduction are exactly the same, as can be seen from the left panel of Fig.~\ref{fig:FVC.mass}. We have also checked that our result of diagram (a) is identical to the one calculated by employing Eq.~\eqref{eq:ruso.mN} (i.e., Eq.~(B27) in Ref.~\cite{Alvarez-Ruso:2013fza}). From Fig.~\ref{fig:FVC.mass}, one could find that the contributions of the nucleon and delta loops are comparable with each other, which implies the importance of the $\Delta$ resonance in the estimation of FVC to the nucleon mass.

In the right panel of Fig.~\ref{fig:FVC.mass}, the $L$ dependence of the FVCs with various values of the pion mass are presented. It can be found that, for a given finite size $L$, the larger the pion mass is, the smaller the FVC become. Fig.~\ref{fig:FVC.mass} also indicates that the effect of FVC on the nucleon mass becomes negligible when $M_\pi L\gtrsim 3$.

\section{Summary and outlook}\label{sec:summary}

A systematical formulation has been advocated to handle the one-loop tensor integrals at finite volume in a universal way. Unlike the infinite case, an extra unit-like space vector arises from the periodical boundary conditions in cubic volume, which makes the decomposition of the tensor integrals cumbersome. We have addressed this problem by incorporating the unit-like space vector in the rank-$P$ tensor, so that a novel decomposition form of the tensor integrals can be established for the calculation of FVC to any physical quantities of interest. As for the tensor coefficients in the decomposition, we have derived a compact formula for them, which is suitable for numerical computations. 

In the CM frame, the tensor coefficients can be simplified to a concise form by means of PV reduction. Two general reduction relations are obtained. When the two formulae are applied recursively, it is found that only the scalar integrals and tensor coefficients proportional to the metric tensors are left to the very end, which can thus be chosen as the tensor basis. 

Lastly, an example is given to illustrate the application of our formulation, where the calculation of the FVC to the nucleon mass is shown step by step. We have also numerically checked that the results with and without PV reduction are exactly the same, indicating the correctness of PV reduction at finite volume.

In summary, our formulation is useful for the calculation of FVC at one-loop level within the framework of perturbation theory, e.g., ChPT. On the other hand, the automation of calculations of FVC often turns out to be challenging due to the necessity of working with quantities that lack Lorentz invariance. The formulation, we have obtained in this work, pave a path for efficient computations of FVC, as it can be easily implemented in high-energy physics packages such as FeynCalc~\cite{Mertig:1990an,Shtabovenko:2016sxi,Shtabovenko:2020gxv} in future.

\acknowledgments
This work is supported by National Nature Science Foundations of China (NSFC) under Contract Nos. 12275076, 11905258 and by the Fundamental Research Funds for the Central Universities under Contract No. 531118010379.


\appendix
\section{Some useful formulae}\label{app.formulae}
The Gaussian parameterization is given in the form as 
\begin{align}
\frac{1}{\mathcal{D}^N}&=\frac{1}{\Gamma(N)}\int^\infty_0 d\lambda \lambda^{N-1}e^{-\lambda \mathcal{D}} \ ,
\label{eq:Gaussian}
\end{align}
with the Gamma function $\Gamma$. The formula for the integration of the $\lambda$ parameter in Eq.~\eqref{eq:Gaussian} is represented by the modified Bessel functions of the second kind $K_z(Y)$~\cite{gradshteyn2014table},
\begin{align}
\mathcal{K}_{z}(X,Y)=\int^\infty_0 d\lambda \lambda^{z-1}e^{-Y\lambda-X/\lambda}
=2\bigg(\frac{X}{Y}\bigg)^{\frac{z}{2}}K_{z}(2\sqrt{XY})
\ .
\label{Eq:Bessel1}
\end{align}
The recurrence relation for the Bessel functions is~\cite{gradshteyn2014table}
\begin{align}
K_{z+1}(Y)-K_{z-1}(Y)=\frac{2z}{Y}K_{z}(Y)
\ .\label{eq:K.rec.rel}
\end{align}

In the Euclidean space, the following kind of momentum integration over $\bar{k}_E$ is needed,
\begin{align}
\int\frac{d^d\bar{k}_E}{(2\pi)^d}(\bar{k}_E^2)^s e^{-\bar{k}_E^2}
&=\frac{1}{(4\pi)^{\frac{d}{2}}}\frac{\Gamma(\frac{d}{2}+s)}{\Gamma(\frac{d}{2})}
=\frac{1}{(4\pi)^{\frac{d}{2}}}\left(\frac{d}{2}\right)_s
\ ,
\label{eq:mom.int}
\end{align}
where $(a)_n$ is the Pochhammer symbol with $n$ being non-negative integers. The definition of the Pochhammer symbol is given by (e.g. see Ref.~\cite{gradshteyn2014table})
\begin{align}
(a)_0=1 \ , \quad 
(a)_1=a \ , \quad 
(a)_{n+1}=a(a+1)\cdots(a+n)=\frac{\Gamma(a+n+1)}{\Gamma(a)}\quad (n\geq1) \ .\label{eq:Pochhammer}
\end{align}

\section{Properties of the $h$-tensor}\label{app.h.tensor}
For completeness, the properties of the auxiliary tensor $h_{\mu\nu}$ in the Minkowski space are collected in this appendix. The $h$-tensor is symmetrical with respect to its Lorentz indices, i.e., $h_{\mu\nu}=h_{\nu\mu}$. In four-dimensional space-time, it reads
$h_{\mu\nu}={\rm diag}(0,-1,-1,-1)$. Contractions of the Lorentz indices of the $h$-tensor lead to the following identities:
\begin{itemize}
\item $h^{\mu}_{\,\mu}=d-1$;
\item $h_{\mu\nu}h^{\mu\nu}=d-1$;
\item $g_{\mu\nu}h^{\mu\nu}=d-1$;
\item $g_{\mu\nu}h^{\mu\rho} h^{\nu\sigma}=h^{\rho\sigma}$;
\item $h_{\mu\nu}p^\mu=(0,-\mathbf{p})$ .
\end{itemize}

\section{Proof of Eq.~\eqref{eq:tensortoscalar}\label{app.proof}}

In a finite box, the rotational symmetry is broken down to the octahedral group $O$. Correspondingly, vectors ${\bf n}$ with integer components transform under the group $O$ as
\bea\label{eq:Otrans}
n_{i}^\prime  =[O^{T_1}(g)]_i^{\,j}n_j\ , \quad g\in O\ , 
\eea
where $O^{T_1}(g)$ is the irreducible representation matrix in a three dimensional space. The matrix $O^{T_1}(g)$ is orthogonal and the transformation keeps the length of the vectors invariant,
\begin{align}\label{eq:inva}
n_s\equiv n_1^2+n_2^2+n_3^2=n_1^{\prime 2}+n_2^{\prime 2}+n_3^{\prime 2}\equiv  n_s^\prime\ .
\end{align}
In consequence, any scalar functions of $n_s$ are invariant. It is also evident that the sum of all the vectors in the whole space remains unchanged, since the effect of transformation under the symmetry group $O$ is merely rotate one vector to another in the same space. Therefore, the following identity holds
\begin{align}
\sum_{{\bf n}\neq0} n_{i_1}\cdots n_{i_{u}}F(-n_s)
&=\sum_{{\bf n}^\prime\neq0} n^\prime_{i_1}\cdots n^\prime_{i_{u}}F(-n_s^{\prime })\ .
\end{align}
With the aid of Eq.~\eqref{eq:Otrans} and Eq.~\eqref{eq:inva}, one obtains
\begin{align}
\sum_{{\bf n}\neq0} n_{i_1}\cdots n_{i_{u}}F(-n_s)
&=\sum_{{\bf n}\neq 0} \left([O^{T_1}(g)]_{i_1}^{\,j_1}n_{j_1}\right)
\cdots
\left([O^{T_1}(g)]_{i_u}^{\,j_u} n_{j_u}\right)F(-n_s)
\nonumber\\
&=[O^{T_1}(g)]_{i_1}^{\,j_1} \cdots [O^{T_1}(g)]_{i_u}^{\,j_u}
\sum_{{\bf n}\neq0} n_{j_1}\cdots n_{j_{u}}F(-n_s)\ .\label{eq:eqind}
\end{align}
For convenience, one may introduce a tensor of rank $u$, 
\begin{align}\label{eq:TT}
\mathcal{T}_{i_1\cdots i_u}\equiv \sum_{{\bf n}\neq0} n_{i_1}\cdots n_{i_{u}}F(-n_s)\ ,
\end{align}
and Eq.~\eqref{eq:eqind} is simplified to
\begin{align}
\mathcal{T}_{i_1i_2\cdots i_u}=[O^{T_1}(g)]_{i_1}^{\,j_1} [O^{T_1}(g)]_{i_2}^{\,j_2}\cdots [O^{T_1}(g)]_{i_u}^{\,j_u}\mathcal{T}_{j_1j_2\cdots j_u}\ ,\quad \forall g\in O\ .
\end{align}
The above equation indicates that the rank-$u$ tensor $\mathcal{T}_{i_1i_2\cdots i_u}$ is an invariant tensor under rotations of group $O$, which is also called isotropic Cartesian tensor~\cite{jeffreys_1973}. Distinct and linearly independent isotropic Cartesian tensors up to rank $6$ are shown in Table~\ref{tab:isotensor}, and the ones of ranks $7$ and $8$ can be found, e.g., in Ref.~\cite{jeffreys_1973, Kearsley1975LinearlyIS,Andrews1981EighthRI}.

\begin{table}[htbp]
\centering
\begin{tabular}{c|c| c}
\hline\hline
Rank  $u$       & Isotropic Cartesian tensors & Total number $N(u)$\\
\hline
$1$ & $-$ & $-$ \\
$2$          &$\delta_{ij}$        & $1$\\
$3$          &$\epsilon_{ijk}$       & $1$ \\
$4$          &$\delta_{ij}\delta_{kl}$\ , \ $\delta_{ik}\delta_{jl}$\ , \  $\delta_{il}\delta_{jk}$ & $3$\\
$5$          &$\epsilon_{ijk}\delta_{lm}$\ , \ $\epsilon_{ijl}\delta_{km}$\ , \ $\epsilon_{ijm}\delta_{kl}$\ , \ $\epsilon_{ikl}\delta_{jm}$\ , \  $\epsilon_{ikm}\delta_{jl}$\ , \ $\epsilon_{ilm}\delta_{jk}$    & $6$\\
$6$          
&$\delta_{ij}\delta_{kl}\delta_{mn}$\ , \  $\delta_{ij}\delta_{km}\delta_{ln}$\ , \  $\delta_{ij}\delta_{kn}\delta_{lm}$\ , \ 
$\delta_{ik}\delta_{jl}\delta_{mn}$\ , \ 
$\delta_{ik}\delta_{jm}\delta_{ln}$\ ,  
& $15$\\
&$\delta_{ik}\delta_{jn}\delta_{lm}$\ , \  $\delta_{il}\delta_{jk}\delta_{mn}$\ , \  
$\delta_{il}\delta_{jm}\delta_{kn}$\ , \  
$\delta_{il}\delta_{jn}\delta_{km}$\ , \ 
$\delta_{im}\delta_{jk}\delta_{ln}$\ ,   
\\
&$\delta_{im}\delta_{jl}\delta_{kn}$\ , \  $\delta_{im}\delta_{jn}\delta_{kl}$\ , \  $\delta_{in}\delta_{jk}\delta_{lm}$\ , \  
$\delta_{in}\delta_{jl}\delta_{km}$\ , \ 
$\delta_{in}\delta_{jm}\delta_{kl}$
\\
\hline\hline
\end{tabular}
\caption{Distinct and independent isotropic Cartesian tensors up to rank $6$. The total number is calculated by using Eq.~\eqref{eq:number}. \label{tab:isotensor}}
\end{table}

In general, as proved in Ref.~\cite{weyl1946classical}, every isotropic Cartesian tensor of even rank can be expressed as a linear combination of products of Kronecker deltas, and every isotropic Cartesian tensor of odd rank is given by a linear combination of products of Kronecker deltas and an alternating tensor. To be specific, the rank-$u$ tensor $\mathcal{T}_{i_1i_2\cdots i_u}$ is written as
\begin{align}\label{eq:isotensor}
\mathcal{T}_{i_1i_2\cdots i_u}
=
\begin{cases}
\sum\limits_{q=1}^{N(u)}\kappa_q \left[\epsilon_{i_{k_1}i_{k_2}i_{k_3}}\delta_{i_{k_4}i_{k_5}}\cdots \delta_{i_{k_{2t}}i_{k_{2t+1}}}\right]_q \ , \quad  u=2t+1 \ , \\
\\
\sum\limits_{p=1}^{N(u)}\lambda_p\, \left[\delta_{i_{k_1}i_{k_2}}\cdots \delta_{i_{k_{2t-1}}i_{k_{2t}}}\right]_p \ ,   \quad  u=2t\ ,
\end{cases}
\end{align}
where $t=1,2,\cdots$ and the $k$'s take values in $\{1,\cdots, u\}$. Furthermore, $N(u)$ is the total number of independent tensors, which can be calculated by the following formulas~\cite{Kearsley1975LinearlyIS},
\begin{align}\label{eq:number}
N(u)=
\begin{cases}
\dfrac{u!}{3!\left(\dfrac{u-3}{2}\right)!2^{(n-3)/2}} \ , \quad  u=2t+1 \ , 
\\
\dfrac{u!}{\left(\dfrac{u}{2}\right)!2^{u/2}} \ , \quad  u=2t \ .
\end{cases}
\end{align}
Here $\kappa_q$ and $\lambda_p$ are unknown coefficients, whose values will be determined below. 

For the case of odd $u$, it can be seen from Eq.~\eqref{eq:isotensor} that, due to the presence of $\epsilon$, there must exist three $i$'s  which are anti-symmetric. However, in view of Eq.~\eqref{eq:TT}, the tensor $\mathcal{T}$ is totally symmetric with respect to its subscripts. Therefore, one may deduce that all the coefficients $\kappa_q$ should be equal to zero. Hence, 
\begin{align}\label{eq:oddcase}
    \mathcal{T}_{i_1i_2\cdots i_{2t+1}}=\sum_{{\bf n}\neq0} n_{i_1}\cdots n_{i_{2t+1}}F(-n_s)=0\ .
\end{align}
It is worth mentioning that the above equation is also true for $u=1$, due to the fact that there is no isotropic Cartesian tensor of rank one, as can be seen from Table~\ref{tab:isotensor}.

For the case of even $u$, thanks to the above mentioned symmetry regarding the indices of the tensor $\mathcal{T}$, the coeffecients $\lambda_p$ are identical. Namely,
\bea
\lambda_1=\lambda_2=\cdots=\lambda_{N(2t)}\equiv \lambda^{(t)}\ ,
\eea
which enables one to rewrite the second line of Eq.~\eqref{eq:isotensor} as 
\begin{align}
\mathcal{T}_{i_1i_2\cdots i_{2t}}=\lambda^{(t)}\, \left\{\delta\cdots \delta\right\}_{i_1i_2\cdots i_{2t}}\ ,
\end{align}
or equivalently
\begin{align}\label{eq:evenid}
\sum_{{\bf n}\neq0} n_{i_1}\cdots n_{i_{2t}}F(-n_s)=\lambda^{(t)}\, \left\{\delta\cdots \delta\right\}_{i_1i_2\cdots i_{2t}}\ .
\end{align}
Note that the notation $\{\cdots\}$ introduced in Eq.~\eqref{eq:p.rank.tensor} has been used. For instance,
\begin{align}
\{\delta\}_{i_1i_2}&=\delta_{i_1i_2}\ ,\\
\{\delta\delta\}_{i_1i_2i_3i_4}&=\delta_{i_1i_2}\delta_{i_3i_4}+\delta_{i_1i_3}\delta_{i_2i_4}+\delta_{i_1i_4}\delta_{i_2i_3}\ ,\\
&\cdots
\end{align}
By contracting $\delta_{i_1i_2}\delta_{i_3i_4}\cdots \delta_{i_{2t-1}i_{2t}}$ on both sides of Eq.~\eqref{eq:evenid}, the coefficient $\lambda^{(t)}$ can be obtained:
\begin{align}
\lambda^{(t)}_p=\frac{1}{(2t+1)!!}\sum_{{\bf n}\neq 0}(n_s)^tF(-n_s)\ .
\end{align}
Substituting it into Eq.~\eqref{eq:evenid}, one gets
\begin{align}\label{eq:evencase}
    \sum_{{\bf n}\neq0} n_{i_1}\cdots n_{i_{2t}}F(-n_s)=\frac{1}{(2t+1)!!}\, \left\{\delta\cdots \delta\right\}_{i_1i_2\cdots i_{2t}}\sum_{{\bf n}\neq 0}(n_s)^tF(-n_s)\ .
\end{align}

In terms of the space vectors $n^\mu=(0,{\bf n})$ and the auxiliary tensor $h_{\mu\nu}$ ($h_{ij}=-\delta_{ij}$ and $h_{\mu\nu}=0$ otherwise), one can recast Eq.~\eqref{eq:oddcase} and Eq.~\eqref{eq:evencase} into forms as
\begin{align}\label{eq:finalform}
\begin{cases}
\sum\limits_{{\bf n}\neq0} n^{\mu_1}\cdots n^{\mu_{2t-1}}F(n^2) =0, \   \text{for odd number of $n^\mu$ }\ , \\
\\
\sum\limits_{{\bf n}\neq0} n^{\mu_1}\cdots n^{\mu_{2t}}F(n^2)=\dfrac{1}{2^t(d_s/2)_t}\, \left\{h\cdots h\right\}^{\mu_1\mu_2\cdots \mu_{2t}}\sum\limits_{{\bf n}\neq 0}(n^2)^tF(n^2),   \ \text{for even number of $n^\mu$}\ ,
\end{cases}
\end{align}
with $d_s=3$ and $(d_s/2)_t$ being the Pochhammer symbol~\eqref{eq:Pochhammer}.

\section{Collection of tensor integrals in arbitrary frame}\label{app.coeff}
In this appendix, explicit expressions for the decomposition of $1$-point, $2$-point, $3$-point and $4$-point tensor integrals up to rank $4$ or $5$ are listed for easy reference.

\subsection{One-point integrals}
The case of $P=0$ corresponds to the scalar integral $\widetilde{A}_0$, which is given by
\begin{align}
\widetilde{A}_0&=\sum_{\mathbf{n}\neq 0}\vecA_0(m^2_1;L,\mathbf{n}) 
\ ,
\end{align}
with 
\begin{align}
\vecA_0(m^2_1;L,\mathbf{n})&=-
\frac{1}{4\pi^2}
\frac{\mathcal{M}_1}{|\mathbf{n}|L}
K_1(|\mathbf{n}|L\mathcal{M}_1)\ ,\quad \mathcal{M}_1=m_1
\ .
\end{align}
The tensor integrals of odd ranks are equal to zero. For rank $2$, the decomposition of the tensor integral $\widetilde{A}^{\mu\nu}$ takes the form
\begin{align}
\widetilde{A}^{\mu\nu}
&=\sum_{\mathbf{n}\neq 0}
\bigg[
g^{\mu\nu}\vecA_{00}
+ 
n^\mu n^\nu 
\vecA_{11}
\bigg]
\ ,
\end{align}
with the coefficients
\begin{align}
\vecA_{00}(m^2_1;L,\mathbf{n})&=
\frac{1}{4\pi^2}
\frac{(\mathcal{M}_1)^2}{|\mathbf{n}|^2L^2}K_2(|\mathbf{n}|L\mathcal{M}_1) \ ,  \\
\vecA_{11}(m^2_1;L,\mathbf{n})&=
\frac{1}{4\pi^2}
\frac{(\mathcal{M}_1)^3}{|\mathbf{n}|^3L}K_3(|\mathbf{n}|L\mathcal{M}_1)  \ .
\end{align}

The one-loop tensor integrals of rank $4$, i.e., $\widetilde{A}^{\mu\nu\rho\sigma}$, can be written as
\begin{align}
\widetilde{A}^{\mu\nu\rho\sigma}
&=\sum_{\mathbf{n}\neq0}
\bigg[
\{g g\}^{\mu\nu\rho\sigma}
\vecA_{0000}
+
\{ g n n \}^{\mu\nu\rho\sigma}
\vecA_{0011}
+
n^\mu n^\nu n^\rho n^\sigma 
\vecA_{1111}
\bigg]
\ ,
\end{align}
with
\begin{align}
\vecA_{0000}(m^2_1;L,\mathbf{n})&=
-\frac{1}{4\pi^2} 
\frac{(\mathcal{M}_1)^3}{|\mathbf{n}|^3L^3}
K_3(|\mathbf{n}|L\mathcal{M}_1)
\ , \\
\vecA_{0011}(m^2_1;L,\mathbf{n})&=
-
\frac{1}{4\pi^2}
\frac{(\mathcal{M}_1)^4}{|\mathbf{n}|^4L^2}
K_4(|\mathbf{n}|L\mathcal{M}_1)
\ , \\
\vecA_{1111}(m^2_1;L,\mathbf{n})&=
-\frac{1}{4\pi^2}
\frac{(\mathcal{M}_1)^5}{|\mathbf{n}|^5L}
K_5(|\mathbf{n}|L\mathcal{M}_1)
\ .
\end{align}

\subsection{Two-point integrals}
The expressions of the two-point tensor integrals up to rank $4$ are explicitly given in the following. The two-point scalar integral $\widetilde{B}_0$ reads 
\begin{align}
\widetilde{B}_0 
&=
\sum_{\mathbf{n}\neq0}
\vecB_{0}(p^2_1,m^2_1,m^2_2;L,\mathbf{n}) 
\ , 
\end{align}
where
\begin{align}
\vecB_{0}(p^2_1,m^2_1,m^2_2;L,\mathbf{n}) 
&=
\frac{1}{8\pi^2}
\int^1_0 \rd X_2
e^{i n\cdot\mathcal{P}_2 L}
K_0(|\mathbf{n}|L\mathcal{M}_2) 
\ ,
\end{align}
and $\int\rd X_2=\int^1_0 \rd x_1$. With the help of the recurrence formulae Eq.~\eqref{eq:mathP} and Eq.~\eqref{eq:mathM}, $\mathcal{P}_2$ and $\mathcal{M}^2_2$ can be obtained, which read
\begin{align}
\mathcal{P}_2&=(1-x_1)p_1\equiv X^1_2p_1 \ , \\
\mathcal{Q}^2_{2}&=x_1 m^2_1+(1-x_1)(m^2_2-p^2_1)
\ , \\
\mathcal{M}^2_2&=\mathcal{Q}^2_{2}+\mathcal{P}^2_{2}
\ .
\end{align}

For tensor integral of rank $1$, 
\begin{align}
\widetilde{B}^{\mu}
&=\sum_{\mathbf{n}\neq0}
\bigg[
p^\mu_1 
\vecB_{1}
+
n^\mu 
\vecB_{2}\bigg]
\ , 
\end{align}
with the coefficients
\begin{align}
\vecB_{1}(p^2_1,m^2_1,m^2_2;L,\mathbf{n}) 
&=
-\frac{1}{8\pi^2}
\int^1_0 \rd X_2 X^1_2
e^{in\cdot{\mathcal{P}}_2 L}
K_0(|\mathbf{n}|L\mathcal{M}_2) 
\ , \\
\vecB_{2}(p^2_1,m^2_1,m^2_2;L,\mathbf{n}) 
&=
\frac{i}{8\pi^2}
\int^1_0 \rd X_2 
e^{i{n}\cdot{\mathcal{P}}_2 L}
\frac{\mathcal{M}_2}{|\mathbf{n}|}
K_1(|\mathbf{n}|L\mathcal{M}_2)
\ .
\end{align}
Here $X^1_2$ is the coefficient of $p_1$ in $\mathcal{P}_2$. It can also be calculated by using the general formula of Eq.~\eqref{eq:mathP2}.

For the case of rank $2$,
\begin{align} 
\widetilde{B}^{\mu\nu}  
&=\sum_{\mathbf{n}\neq0}
\bigg[
g^{\mu\nu}
\vecB_{00}
+
p^\mu_1 p^\nu_1 
\vecB_{11}
+
\{ p_1 n \}^{\mu\nu}
\vecB_{12}
+
n^\mu n^\nu 
\vecB_{22}
\bigg]
\ , 
\end{align}
with the corresponding coefficients
\begin{align}
\vecB_{00}(p^2_1,m^2_1,m^2_2;L,\mathbf{n}) 
&=
-
\frac{1}{8\pi^2}
\int^1_0 \rd X_2
e^{i{n}\cdot{\mathcal{P}}_2 L}
\frac{\mathcal{M}_2}{|\mathbf{n}|L}
K_1(|\mathbf{n}|L\mathcal{M}_2) \ , \\
\vecB_{11}(p^2_1,m^2_1,m^2_2;L,\mathbf{n}) 
&=
\frac{1}{8\pi^2}
\int^1_0 \rd X_2 (X^1_2)^2
e^{i{n}\cdot{\mathcal{P}}_2 L}
K_0(|\mathbf{n}|L\mathcal{M}_2) \ , \\
\vecB_{12}(p^2_1,m^2_1,m^2_2;L,\mathbf{n}) 
&=
-\frac{i}{8\pi^2}
\int^1_0\rd X_2 X^1_2
e^{i{n}\cdot{\mathcal{P}}_2 L}
\frac{\mathcal{M}_2}{|\mathbf{n}|}
K_1(|\mathbf{n}|L\mathcal{M}_2)
\ , \\
\vecB_{22}(p^2_1,m^2_1,m^2_2;L,\mathbf{n}) 
&=
-
\frac{1}{8\pi^2}
\int^1_0\rd X_2
e^{i{n}\cdot{\mathcal{P}}_2 L}
\frac{(\mathcal{M}_2)^2}{|\mathbf{n}|^2}
K_2(|\mathbf{n}|L\mathcal{M}_2) \ .
\end{align}

For rank $3$, one has
\begin{align}
\widetilde{B}^{\mu\nu\rho}
&=
\sum_{\mathbf{n}\neq0}\bigg[ 
\{ g p_1 \}^{\mu\nu\rho}
\vecB_{001} 
+
\{ g n \}^{\mu\nu\rho}
\vecB_{002}  
+
p^\mu_1 p^\nu_1 p^\rho_1 
\vecB_{111}
\notag \\
&
+
\{ p_1 p_1 n \}^{\mu\nu\rho} 
\vecB_{112} 
+
\{ p_1 n n \}^{\mu\nu\rho}
\vecB_{122} 
+
n^\mu n^\nu n^\rho 
\vecB_{222} 
\bigg]
\ ,
\end{align}
with the coefficients
\begin{align}
\vecB_{001}(p^2_1,m^2_1,m^2_2;L,\mathbf{n})
&=
\frac{1}{8\pi^2}
\int^1_0\rd X_2 X^1_2
e^{i{n}\cdot{\mathcal{P}}_2 L} 
\frac{\mathcal{M}_2}{|\mathbf{n}|L}
K_1(|\mathbf{n}|L\mathcal{M}_2)  
\ , \\
\vecB_{002}(p^2_1,m^2_1,m^2_2;L,\mathbf{n})
&=
-\frac{i}{8\pi^2}
\int^1_0 \rd X_2
e^{i{n}\cdot{\mathcal{P}}_2 L} 
\frac{(\mathcal{M}_2)^2}{|\mathbf{n}|^2L}
K_2(|\mathbf{n}|L\mathcal{M}_2)
\ , \\
\vecB_{111}(p^2_1,m^2_1,m^2_2;L,\mathbf{n}) 
&=
-
\frac{1}{8\pi^2}
\int^1_0\rd X_2 (X^1_2)^3
e^{i{n}\cdot{\mathcal{P}}_2 L} 
K_0(|\mathbf{n}|L\mathcal{M}_2)
\ , \\
\vecB_{112}(p^2_1,m^2_1,m^2_2;L,\mathbf{n}) 
&=
\frac{i}{8\pi^2}
\int^1_0\rd X_2(X^1_2)^2
e^{i{n}\cdot{\mathcal{P}}_2 L} 
\frac{\mathcal{M}_2}{|\mathbf{n}|}
K_1(|\mathbf{n}|L\mathcal{M}_2)
\ , \\
\vecB_{122}(p^2_1,m^2_1,m^2_2;L,\mathbf{n}) 
&=
\frac{1}{8\pi^2}
\int^1_0\rd X_2 X^1_2
e^{i{n}\cdot{\mathcal{P}}_2 L} 
\frac{(\mathcal{M}_2)^2}{|\mathbf{n}|^2}
K_2(|\mathbf{n}|L\mathcal{M}_2)
\ , \\
\vecB_{222}(p^2_1,m^2_1,m^2_2;L,\mathbf{n}) 
&=
-
\frac{i}{8\pi^2}
\int^1_0\rd X_2
e^{i{n}\cdot{\mathcal{P}}_2 L} 
\frac{(\mathcal{M}_2)^3}{|\mathbf{n}|^3}
K_3(|\mathbf{n}|L\mathcal{M}_2)
\ .
\end{align}

For the tensor integral of rank $4$, we have 
\begin{align}
\widetilde{B}^{\mu\nu\rho\sigma}&=\sum_{\mathbf{n}\neq0}
\bigg[
\{g g\}^{\mu\nu\rho\sigma}
\vecB_{0000}
+\{g p_1 p_1\}^{\mu\nu\rho\sigma}
\vecB_{0011}
+\{g p_1 n\}^{\mu\nu\rho\sigma}
\vecB_{0012}
+\{g n n\}^{\mu\nu\rho\sigma}
\vecB_{0022}
\notag \\
&+p^\mu_1 p^\nu_1 p^\rho_1 p^\sigma_1\vecB_{1111}
+\{p_1 p_1 p_1 n\}^{\mu\nu\rho\sigma}
\vecB_{1112}
+\{ p_1 p_1 n n\}^{\mu\nu\rho\sigma}
\vecB_{1122}
+\{ p_1 n n n\}^{\mu\nu\rho\sigma}
\vecB_{1222}
\notag \\
&
+n^\mu n^\nu n^\rho n^\sigma
\vecB_{2222}
\bigg]\ ,
\end{align}
with the coefficients 
\begin{align}
\vecB_{0000}(p^2_1,m^2_1,m^2_2;L,\mathbf{n}) &=\frac{1}{8\pi^2}\int^1_0\rd X_2
e^{i{n}\cdot{\mathcal{P}}_2 L}
\frac{(\mathcal{M}_2)^2}{|\mathbf{n}|^2L^2}
K_2(|\mathbf{n}|L\mathcal{M}_2)
\ , \\
\vecB_{0011}(p^2_1,m^2_1,m^2_2;L,\mathbf{n}) &=-\frac{1}{8\pi^2}\int^1_0\rd X_2 (X^1_2)^2
e^{i{n}\cdot{\mathcal{P}}_2 L}
\frac{\mathcal{M}_2}{|\mathbf{n}|L}
K_1(|\mathbf{n}|L\mathcal{M}_2)
\ , \\
\vecB_{0012}(p^2_1,m^2_1,m^2_2;L,\mathbf{n})
&=\frac{i}{8\pi^2}\int^1_0\rd X_2 X^1_2
e^{i{n}\cdot{\mathcal{P}}_2 L}
\frac{(\mathcal{M}_2)^2}{|\mathbf{n}|^2L}
K_2(|\mathbf{n}|L\mathcal{M}_2)
\ , \\
\vecB_{0022}(p^2_1,m^2_1,m^2_2;L,\mathbf{n})&=\frac{1}{8\pi^2}\int^1_0\rd X_2 
e^{i{n}\cdot{\mathcal{P}}_2 L}
\frac{(\mathcal{M}_2)^3}{|\mathbf{n}|^3L}
K_3(|\mathbf{n}|L\mathcal{M}_2)
\ , \\
\vecB_{1111}(p^2_1,m^2_1,m^2_2;L,\mathbf{n}) &=\frac{1}{8\pi^2}\int^1_0\rd X_2
(X^1_2)^4
e^{i{n}\cdot{\mathcal{P}}_2 L} K_0(|\mathbf{n}|L\mathcal{M}_2)
\ , \\
\vecB_{1112}(p^2_1,m^2_1,m^2_2;L,\mathbf{n}) &=-\frac{i}{8\pi^2}\int^1_0\rd X_2 (X^1_2)^3
e^{i{n}\cdot{\mathcal{P}}_2 L}
\frac{\mathcal{M}_2}{|\mathbf{n}|}
K_1(|\mathbf{n}|L\mathcal{M}_2)
\ , \\
\vecB_{1122}(p^2_1,m^2_1,m^2_2;L,\mathbf{n}) &=-\frac{1}{8\pi^2}\int^1_0\rd X_2 (X^1_2)^2
e^{i{n}\cdot{\mathcal{P}}_2 L}
\frac{(\mathcal{M}_2)^2}{|\mathbf{n}|^2}
K_2(|\mathbf{n}|L\mathcal{M}_2)
\ , \\
\vecB_{1222}(p^2_1,m^2_1,m^2_2;L,\mathbf{n}) &=\frac{i}{8\pi^2}\int^1_0\rd X_2 X^1_2
e^{i{n}\cdot{\mathcal{P}}_2 L}
\frac{(\mathcal{M}_2)^3}{|\mathbf{n}|^3}
K_3(|\mathbf{n}|L\mathcal{M}_2)
\ , \\
\vecB_{2222}(p^2_1,m^2_1,m^2_2;L,\mathbf{n}) &=\frac{1}{8\pi^2}\int^1_0\rd X_2
e^{i{n}\cdot{\mathcal{P}}_2 L}
\frac{(\mathcal{M}_2)^4}{|\mathbf{n}|^4}
K_4(|\mathbf{n}|L\mathcal{M}_2)
\ .
\end{align}

\subsection{Three-point integrals}
The decomposition of the three-point integrals up to rank $4$ and the involved coefficients are given in the following. The three-point scalar integral $\widetilde{C}_0$ reads
\begin{align}
\widetilde{C}_0
&=
\sum_{\mathbf{n}\neq0}
\vecC_0(p_1^2, (p_2-p_1)^2, p_2^2, m^2_1, m^2_2, m^2_3;L,\mathbf{n})
\ ,    
\end{align}
with  
\begin{align}
\vecC_0(p_1^2, (p_2-p_1)^2, p_2^2, m^2_1, m^2_2, m^2_3;L,\mathbf{n})&=-\frac{1}{16\pi^2}\int^1_0 \rd X_3
e^{i{n}\cdot{\mathcal{P}}_3 L}
\frac{|\mathbf{n}|L}{\mathcal{M}_3}K_1(|\mathbf{n}|L\mathcal{M}_3)
\ ,
\end{align}
and $\int \rd X_3=\int^1_0 \rd x_1\int^1_0 \rd x_2 x_2$. Hereafter, the arguments of the $\vecC$ functions will be suppressed for brevity. The $\mathcal{P}_3$ and $\mathcal{M}^2_3$ can be obtained by the recurrence formulae Eq.~\eqref{eq:mathP} and Eq.~\eqref{eq:mathM}, which read
\begin{align}
\mathcal{P}_3&=x_2(1-x_1)p_1+(1-x_2)p_2\equiv\sum^2_{i=1}X^i_3 p_i \ ,\label{eq:MP3} \\
\mathcal{Q}^2_{3}&=x_2\mathcal{Q}^2_{2}+(1-x_2)(m^2_3-p^2_2) \ , \\
\mathcal{M}^2_3&=\mathcal{Q}^2_{3}+\mathcal{P}^2_{3}  \ .
\end{align}

For the tensor integral of rank $1$, the decomposition is 
\begin{align}
\widetilde{C}^{\mu}
&=
\sum_{\mathbf{n}\neq0}
\bigg[ 
\sum^2_{i=1} p_i^\mu 
\vecC_{i}
+
n^\mu \vecC_3
\bigg] 
\ ,    
\end{align}
where the coefficients are
\begin{align}
\vecC_{i}
&=
\frac{1}{16\pi^2}
\int^1_0 \rd X_3 X^i_3
e^{i{n}\cdot{\mathcal{P}}_3 L}
\frac{|\mathbf{n}|L}{\mathcal{M}_3}
K_{1}(|\mathbf{n}|L\mathcal{M}_3) \ ,
\\
\vecC_3
&=
-\frac{i L}{16\pi^2}
\int^1_0 \rd X_3 
e^{i{n}\cdot{\mathcal{P}}_3 L}
K_{0}(|\mathbf{n}|L\mathcal{M}_3)
\ .
\end{align}
Here, $X^1_3$ and $X^2_3$ are the coefficients of $p_1$ and $p_2$ in $\mathcal{P}_3$~\eqref{eq:MP3}, respectively.

For rank $2$, one has
\begin{align}
\widetilde{C}^{\mu\nu}
&=
\sum_{\mathbf{n}\neq0}
\bigg[ 
g^{\mu\nu} 
\vecC_{00}
+
\sum^2_{i,j=1}p^\mu_i p^\nu_j 
\vecC_{ij}
+
\sum^2_{i=1}
\{ p n \}^{\mu\nu}_{i} 
\vecC_{i3}
+
n^\mu n^\nu 
\vecC_{33}
\bigg] 
\ , 
\end{align}
with the coefficients
\begin{align}
\vecC_{00}
&=
\frac{1}{16\pi^2}
\int^1_0 \rd X_3
e^{i{n}\cdot{\mathcal{P}}_3 L}
K_0(|\mathbf{n}|L\mathcal{M}_3) 
\ , \\
\vecC_{ij}
&=
-\frac{1}{16\pi^2}
\int^1_0 \rd X_3
~X^i_3 X^j_3
e^{i{n}\cdot{\mathcal{P}}_3 L}
\frac{|\mathbf{n}|L}{\mathcal{M}_3}
K_1(|\mathbf{n}|L\mathcal{M}_3)  
\ , \\
\vecC_{i3}
&=
\frac{iL}{16\pi^2}
\int^1_0 \rd X_3
~X^i_3
e^{i{n}\cdot{\mathcal{P}}_3 L}
K_0(|\mathbf{n}|L\mathcal{M}_3) 
\ , \\
\vecC_{33}
&=
\frac{1}{16\pi^2}
\int^1_0 \rd X_3
e^{i{n}\cdot{\mathcal{P}}_3 L}
\frac{\mathcal{M}_3L}{|\mathbf{n}|}
K_1(|\mathbf{n}|L\mathcal{M}_3)  \ .
\end{align}

For the case of rank $3$, 
\begin{align}
\widetilde{C}^{\mu\nu\rho}
&=
\sum_{\mathbf{n}\neq0}
\bigg[ 
\sum^2_{i=1}
\{ g p \}^{\mu\nu\rho}_{i}
\vecC_{00i}
+
\{ g n \}^{\mu\nu\rho} 
\vecC_{003}
+
\sum^2_{i,j,k=1}
p^\mu_i p^\nu_j p^\rho_k
\vecC_{ijk}
+
\sum^2_{i,j=1}
\{ p p n \}^{\mu\nu\rho}_{ij}
\vecC_{ij3}
\notag \\
&
+
\sum^2_{i=1}
\{ p n n \}^{\mu\nu\rho}_{i}
\vecC_{i33}
+
\{ n n n \}^{\mu\nu\rho}
\vecC_{333}
\bigg] 
\ , 
\end{align}
with
\begin{align}
\vecC_{00i}&=
-\frac{1}{16\pi^2}\int^1_0 \rd X_3 X^i_3
e^{i{n}\cdot{\mathcal{P}}_3 L}
K_0(|\mathbf{n}|L\mathcal{M}_3) 
\ , \\
\vecC_{003}&=
\frac{i}{16\pi^2}\int^1_0 \rd X_3
e^{i{n}\cdot{\mathcal{P}}_3 L}
\frac{\mathcal{M}_3}{|\mathbf{n}|}K_1(|\mathbf{n}|L\mathcal{M}_3) 
\ , \\
\vecC_{ijk}&=
\frac{1}{16\pi^2}\int^1_0 \rd X_3 X^i_3 X^j_3 X^k_3
e^{i{n}\cdot{\mathcal{P}}_3 L}
\frac{|\mathbf{n}|L}{\mathcal{M}_3}K_1(|\mathbf{n}|L\mathcal{M}_3) 
\ , \\
\vecC_{ij3}&=
-\frac{iL}{16\pi^2}\int^1_0 \rd X_3 
X^i_3 X^j_3
e^{i{n}\cdot{\mathcal{P}}_3 L}
K_0(|\mathbf{n}|L\mathcal{M}_3) 
\ , \\
\vecC_{i33}&=
-\frac{1}{16\pi^2}\int^1_0 \rd X_3 
X^i_3
e^{i{n}\cdot{\mathcal{P}}_3 L}
\frac{\mathcal{M}_3 L}{|\mathbf{n}|}
K_1(|\mathbf{n}|L\mathcal{M}_3) 
\ , \\
\vecC_{333}&=
\frac{i}{16\pi^2}\int^1_0 \rd X_3
e^{i{n}\cdot{\mathcal{P}}_3 L}
\frac{(\mathcal{M}_3)^2L}{|\mathbf{n}|^2}
K_2(|\mathbf{n}|L\mathcal{M}_3)
\ .
\end{align}

For the case of rank $4$, 
\begin{align}
\widetilde{C}^{\mu\nu\rho\sigma}
&=
\sum_{\mathbf{n}\neq0}
\bigg[ 
\{ g g \}^{\mu\nu\rho\sigma}
\vecC_{0000}
+
\sum^2_{i,j=1}\{ g p p \}^{\mu\nu\rho\sigma}_{ij}
\vecC_{00ij}
+
\sum^2_{i=1} \{ g p n \}^{\mu\nu\rho\sigma}_i
\vecC_{00i3}
\notag \\
&
+
\{ g n n \}^{\mu\nu\rho\sigma}
\vecC_{0033}
+
\sum^2_{i,j,k,l=1}
p^\mu_i p^\nu_j p^\rho_k p^\sigma_l
\vecC_{ijkl}
+
\sum^2_{i,j,k=1}\{ p p p n \}^{\mu\nu\rho\sigma}_{ijk}
\vecC_{ijk3}
\notag \\
&
+
\sum^2_{i,j=1}\{ p p n n \}^{\mu\nu\rho\sigma}_{ij}
\vecC_{ij33}
+
\sum^2_{i=1} \{ p n n n \}^{\mu\nu\rho\sigma}_i
\vecC_{i333}
+
n^\mu n^\nu n^\rho n^\sigma 
\vecC_{3333}
\bigg] 
\ ,
\end{align}
with
\begin{align}
\vecC_{0000}&=
-\frac{1}{16\pi^2}\int^1_0 \rd X_3 
e^{i{n}\cdot{\mathcal{P}}_3 L}
\frac{\mathcal{M}_3}{|\mathbf{n}|L}
K_1(|\mathbf{n}|L\mathcal{M}_3)
\ , \\
\vecC_{00ij}&=
\frac{1}{16\pi^2}\int^1_0 \rd X_3 
X^i_3 X^j_3
e^{i{n}\cdot{\mathcal{P}}_3 L}
K_0(|\mathbf{n}|L\mathcal{M}_3)
\ , \\
\vecC_{00i3}&=
-\frac{i}{16\pi^2}\int^1_0 \rd X_3 
X^i_3
e^{i{n}\cdot{\mathcal{P}}_3 L}
\frac{\mathcal{M}_3}{|\mathbf{n}|}
K_1(|\mathbf{n}|L\mathcal{M}_3)
\ , \\
\vecC_{0033}&=
-\frac{1}{16\pi^2}\int^1_0 \rd X_3 
e^{i{n}\cdot{\mathcal{P}}_3 L}
\frac{(\mathcal{M}_3)^2}{|\mathbf{n}|^2}
K_2(|\mathbf{n}|L\mathcal{M}_3)
\ , \\
\vecC_{ijkl}&=
-\frac{1}{16\pi^2}\int^1_0 \rd X_3 
X^i_3 X^j_3 X^k_3 X^l_3
e^{i{n}\cdot{\mathcal{P}}_3 L}
\frac{|\mathbf{n}|L}{\mathcal{M}_3}
K_1(|\mathbf{n}|L\mathcal{M}_3)
\ , \\
\vecC_{ijk3}&=
\frac{iL}{16\pi^2}\int^1_0 \rd X_3 
X^i_3 X^j_3 X^k_3
e^{i{n}\cdot{\mathcal{P}}_3 L}
K_0(|\mathbf{n}|L\mathcal{M}_3)
\ , \\
\vecC_{ij33}&=
\frac{1}{16\pi^2}\int^1_0 \rd X_3 
X^i_3 X^j_3
e^{i{n}\cdot{\mathcal{P}}_3 L}
\frac{\mathcal{M}_3L}{|\mathbf{n}|}
K_1(|\mathbf{n}|L\mathcal{M}_3)
\ , \\
\vecC_{i333}&=
-\frac{i}{16\pi^2}\int^1_0 \rd X_3 
X^i_3
e^{i{n}\cdot{\mathcal{P}}_3 L}
\frac{(\mathcal{M}_3)^2L}{|\mathbf{n}|^2}
K_2(|\mathbf{n}|L\mathcal{M}_3)
\ , \\
\vecC_{3333}&=
-\frac{1}{16\pi^2}\int^1_0 \rd X_3 
e^{i{n}\cdot{\mathcal{P}}_3 L}
\frac{(\mathcal{M}_3)^3L}{|\mathbf{n}|^3}
K_3(|\mathbf{n}|L\mathcal{M}_3)
\ .
\end{align}

\subsection{Four-point integrals} 
The decomposition of the four-point integrals up to rank $5$ and the involved coefficients are given in the following. The four-point scalar integral $\widetilde{D}_0$ reads 
\begin{align}
\widetilde{D}_0
&=\sum_{\mathbf{n}\neq0}\vecD_0(p_1^2, (p_2-p_1)^2, (p_3-p_2)^2, p_3^2, p_2^2, (p_3-p_1)^2, m^2_1, m^2_2, m^2_3, m^2_4; L, \mathbf{n})
\ ,
\end{align}
with
\begin{align}
\vecD_0&=
\frac{1}{32\pi^2}\int^1_0 \rd X_4
e^{i{n}\cdot{\mathcal{P}}_4 L}
\frac{|\mathbf{n}|^2L^2}{(\mathcal{M}_4)^2}
K_2(|\mathbf{n}|L\mathcal{M}_4)
\ ,
\end{align}
and $\int\rd X_4=\int^1_0 \rd x_1\int^1_0 \rd x_2 \int^1_0 \rd x_3 x_2 x^2_3$. Hereafter, the arguments of the $\vecD$ functions will be suppressed for brevity. The $\mathcal{P}_4$ and $\mathcal{M}^2_4$ can be obtained by the recurrence formulae Eq.~\eqref{eq:mathP} and Eq.~\eqref{eq:mathM}, which read
\begin{align}
\mathcal{P}_4&=
x_3x_2(1-x_1)p_1+x_3(1-x_2)p_2+(1-x_3)p_3\equiv\sum^3_{i=1}X^i_4 p_i
\ , \label{eq:MP4} \\
\mathcal{Q}^2_{4}&=x_3\mathcal{Q}^2_{3}+(1-x_3)(m^2_4-p^2_3)
\ , \\
\mathcal{M}^2_4&=\mathcal{Q}^2_{4}+\mathcal{P}^2_{4} \ .
\end{align}

For the tensor integral of rank $1$, the decomposition is 
\begin{align}
\widetilde{D}^{\mu}
&=
\sum_{\mathbf{n}\neq0}
\bigg[ 
\sum^3_{i=1} p^\mu_i 
\vecD_{i}
+
n^\mu \vecD_4
\bigg] 
\ ,
\end{align}
where the coefficients are
\begin{align}
\vecD_i&=
-\frac{1}{32\pi^2}\int^1_0 \rd X_4
X^i_4
e^{i{n}\cdot{\mathcal{P}}_4 L}
\frac{|\mathbf{n}|^2L^2}{(\mathcal{M}_4)^2}
K_2(|\mathbf{n}|L\mathcal{M}_4)
\ , \\
\vecD_4&=
\frac{i}{32\pi^2}\int^1_0 \rd X_4
e^{i{n}\cdot{\mathcal{P}}_4 L}
\frac{|\mathbf{n}|L^2}{\mathcal{M}_4}
K_1(|\mathbf{n}|L\mathcal{M}_4)
\ .
\end{align}
Here, $X^1_4$, $X^2_4$ and $X^3_4$ are the coefficients of $p_1$, $p_2$ and $p_3$ in $\mathcal{P}_4$~\eqref{eq:MP4}, in order.

For rank $2$, one obtains
\begin{align}
\widetilde{D}^{\mu\nu}
&=
\sum_{\mathbf{n}\neq0}
\bigg[ 
g^{\mu\nu} 
\vecD_{00}
+
\sum^3_{i,j=1}
p^\mu_i p^\nu_j 
\vecD_{ij}
+
\sum^3_{i=1} \{ p n \}^{\mu\nu}_i
\vecD_{i4}
+
n^\mu n^\nu 
\vecD_{44}
\bigg] 
\ ,
\end{align}
with the coefficients 
\begin{align}
\vecD_{00}&=
-\frac{1}{32\pi^2}\int^1_0 \rd X_4
e^{i{n}\cdot{\mathcal{P}}_4 L}
\frac{|\mathbf{n}|L}{\mathcal{M}_4}
K_1(|\mathbf{n}|L\mathcal{M}_4)
\ , \\
\vecD_{ij}&=
\frac{1}{32\pi^2}\int^1_0 \rd X_4
X^i_4 X^j_4
e^{i{n}\cdot{\mathcal{P}}_4 L}
\frac{|\mathbf{n}|^2L^2}{(\mathcal{M}_4)^2}
K_2(|\mathbf{n}|L\mathcal{M}_4)
\ , \\
\vecD_{i4}&=
-\frac{i}{32\pi^2}\int^1_0 \rd X_4
X^i_4
e^{i{n}\cdot{\mathcal{P}}_4 L}
\frac{|\mathbf{n}|L^2}{\mathcal{M}_4}
K_1(|\mathbf{n}|L\mathcal{M}_4)
\ , \\
\vecD_{44}&=
-\frac{L^2}{32\pi^2}\int^1_0 \rd X_4
e^{i{n}\cdot{\mathcal{P}}_4 L}
K_0(|\mathbf{n}|L\mathcal{M}_4)
\ .
\end{align}

For rank $3$, one has
\begin{align}
\widetilde{D}^{\mu\nu\rho}
&=
\sum_{\mathbf{n}\neq0}
\bigg[ 
\sum^3_{i=1}\{ g p \}^{\mu\nu\rho}_i
\vecD_{00i}
+
\{ g n \}^{\mu\nu\rho}
\vecD_{004}
+
\sum^3_{i,j,k=1}p^\mu_i p^\nu_j p^\rho_k \vecD_{ijk}
+
\sum^3_{i,j=1} 
\{ p p n \}^{\mu\nu\rho}_{ij}
\vecD_{ij4}
\notag \\
&
+
\sum^3_{i=1}
\{ p n n \}^{\mu\nu\rho}_i
\vecD_{i44}
+
n^\mu n^\nu n^\rho 
\vecD_{444}
\bigg] 
\ ,
\end{align}
with
\begin{align}
\vecD_{00i}&=
\frac{1}{32\pi^2}\int^1_0 \rd X_4
X^i_4
e^{i{n}\cdot{\mathcal{P}}_4 L}
\frac{|\mathbf{n}|L}{\mathcal{M}_4}
K_1(|\mathbf{n}|L\mathcal{M}_4)
\ , \\
\vecD_{004}&=
-\frac{iL}{32\pi^2}\int^1_0 \rd X_4
e^{i{n}\cdot{\mathcal{P}}_4 L}
K_0(|\mathbf{n}|L\mathcal{M}_4)
\ , \\
\vecD_{ijk}&=
-\frac{1}{32\pi^2}\int^1_0 \rd X_4
X^i_4 X^j_4 X^k_4
e^{i{n}\cdot{\mathcal{P}}_4 L}
\frac{|\mathbf{n}|^2L^2}{(\mathcal{M}_4)^2}
K_2(|\mathbf{n}|L\mathcal{M}_4)
\ , \\
\vecD_{ij4}&=
\frac{i}{32\pi^2}\int^1_0 \rd X_4
X^i_4 X^{j}_4
e^{i{n}\cdot{\mathcal{P}}_4 L}
\frac{|\mathbf{n}|L^2}{\mathcal{M}_4}
K_1(|\mathbf{n}|L\mathcal{M}_4)
\ , \\
\vecD_{i44}&=
\frac{L^2}{32\pi^2}\int^1_0 \rd X_4
X^i_4
e^{i{n}\cdot{\mathcal{P}}_4 L}
K_0(|\mathbf{n}|L\mathcal{M}_4)
\ , \\
\vecD_{444}&=
-\frac{i}{32\pi^2}\int^1_0 \rd X_4
e^{i{n}\cdot{\mathcal{P}}_4 L}
\frac{L^2\mathcal{M}_4}{|\mathbf{n}|}
K_1(|\mathbf{n}|L\mathcal{M}_4)
\ .
\end{align}

For rank $4$, we have
\begin{align}
\widetilde{D}^{\mu\nu\rho\sigma}
&=
\sum_{\mathbf{n}\neq0}
\bigg[ 
\{ g g \}^{\mu\nu\rho\sigma}
\vecD_{0000}
+
\sum^3_{i,j=1}\{ g p p \}^{\mu\nu\rho\sigma}_{ij}
\vecD_{00ij}
+\sum^{3}_{i=1}\{ g p n\}^{\mu\nu\rho\sigma}
\vecD_{00i4}
\notag \\
&
+\{g n n\}^{\mu\nu\rho\sigma}
\vecD_{0044}
+
\sum^3_{i,j,k,l=1}p^\mu_i p^\nu_j p^\rho_k p^\sigma_l \vecD_{ijkl}
+
\sum^3_{i,j,k=1}\{ p p p n \}^{\mu\nu\rho\sigma}_{ijk}
\vecD_{ijk4}
\notag \\
&
+
\sum^3_{i,j=1}\{ p p n n \}^{\mu\nu\rho\sigma}_{ij}
\vecD_{ij44}
+
\sum^3_{i=1} \{ p n n n \}^{\mu\nu\rho\sigma}_i
\vecD_{i444}
+n^\mu n^\nu n^\rho n^\sigma 
\vecD_{4444}
\bigg] 
\ , 
\end{align}
with the coefficients
\begin{align}
\vecD_{0000}&=
\frac{1}{32\pi^2}\int^1_0 \rd X_4
e^{i{n}\cdot{\mathcal{P}}_4 L}
K_0(|\mathbf{n}|L\mathcal{M}_4)
\ , \\
\vecD_{00ij}&=
-\frac{1}{32\pi^2}\int^1_0 \rd X_4
X^i_4 X^j_4
e^{i{n}\cdot{\mathcal{P}}_4 L}
\frac{|\mathbf{n}|L}{\mathcal{M}_4}
K_1(|\mathbf{n}|L\mathcal{M}_4)
\ , \\
\vecD_{00i4}&=
\frac{iL}{32\pi^2}
\int^1_0 \rd X_4 X^i_4
e^{i{n}\cdot{\mathcal{P}}_4 L}
K_0(|\mathbf{n}|L\mathcal{M}_4)
\ , \\
\vecD_{0044}&=
\frac{1}{32\pi^2}
\int^1_0 \rd X_4
e^{i{n}\cdot{\mathcal{P}}_4 L}
\frac{\mathcal{M}_4L}{|\mathbf{n}|}
K_1(|\mathbf{n}|L\mathcal{M}_4)
\ , \\
\vecD_{ijkl}&=
\frac{1}{32\pi^2}\int^1_0 \rd X_4
X^i_4 X^j_4
X^k_4 X^l_4
e^{i{n}\cdot{\mathcal{P}}_4 L}
\frac{|\mathbf{n}|^2L^2}{(\mathcal{M}_4)^2}
K_2(|\mathbf{n}|L\mathcal{M}_4)
\ , \\
\vecD_{ijk4}&=
-\frac{i}{32\pi^2}\int^1_0 \rd X_4
X^i_4 X^j_4
X^k_4
e^{i{n}\cdot{\mathcal{P}}_4 L}
\frac{|\mathbf{n}|L^2}{\mathcal{M}_4}
K_1(|\mathbf{n}|L\mathcal{M}_4)
\ , \\
\vecD_{ij44}&=
-\frac{L^2}{32\pi^2}\int^1_0 \rd X_4
X^i_4 X^j_4
e^{i{n}\cdot{\mathcal{P}}_4 L}
K_0(|\mathbf{n}|L\mathcal{M}_4)
\ , \\
\vecD_{i444}&=
\frac{i}{32\pi^2}\int^1_0 \rd X_4
X^i_4
e^{i{n}\cdot{\mathcal{P}}_4 L}
\frac{L^2\mathcal{M}_4}{|\mathbf{n}|}
K_1(|\mathbf{n}|L\mathcal{M}_4)
\ , \\
\vecD_{4444}&=
\frac{1}{32\pi^2}\int^1_0 \rd X_4
e^{i{n}\cdot{\mathcal{P}}_4 L}
\frac{L^2(\mathcal{M}_4)^2}{|\mathbf{n}|^2}
K_2(|\mathbf{n}|L\mathcal{M}_4)
\ .
\end{align}

For the case of rank $5$, the decomposition is given by
\begin{align}
\widetilde{D}^{\mu\nu\rho\sigma\alpha}
&=
\sum_{\mathbf{n}\neq0}
\bigg[ 
\sum^3_{i=1}
\{ g g p \}^{\mu\nu\rho\sigma\alpha}_{i}
\vecD_{0000i}
+\{ g g n \}^{\mu\nu\rho\sigma\alpha}
\vecD_{00004}
+
\sum^3_{i,j,k=1}
\{ g p p p \}^{\mu\nu\rho\sigma\alpha}_{ijk}
\vecD_{00ijk}
\notag \\
&
+\sum^3_{i,j=1}
\{ g p p n \}^{\mu\nu\rho\sigma\alpha}_{ij}
\vecD_{00ij4}
+\sum^3_{i=1}
\{ g p n n \}^{\mu\nu\rho\sigma\alpha}_{i}
\vecD_{00i44}
+\{ g n n n \}^{\mu\nu\rho\sigma\alpha}
\vecD_{00444}
\notag \\
&
+\sum^3_{i,j,k,l,r=1}
p^\mu_i p^\nu_j p^\rho_k p^\sigma_l p^\alpha_r 
\vecD_{ijklr}
+\sum^3_{i,j,k,l=1}
\{ p p p p n \}^{\mu\nu\rho\sigma\alpha}_{ijkl}
\vecD_{ijkl4}
+
\sum^3_{i,j,k=1}
\{ p p p n n \}^{\mu\nu\rho\sigma\alpha}_{ijk}
\vecD_{ijk44}
\notag \\
&
+
\sum^3_{i,j=1}
\{ p p n n n \}^{\mu\nu\rho\sigma\alpha}_{ij}
\vecD_{ij444}
+
\sum^3_{i=1}
\{ p n n n n \}^{\mu\nu\rho\sigma\alpha}_i
\vecD_{i4444}
+
n^\mu n^\nu n^\rho n^\sigma n^\alpha 
\vecD_{44444}
\bigg] 
\ ,
\end{align}
with the coefficients
\begin{align}
\vecD_{0000i}&=
-\frac{1}{32\pi^2}
\int^1_0 \rd X_4 X^i_4
e^{i{n}\cdot{\mathcal{P}}_4 L}
K_0(|\mathbf{n}|L\mathcal{M}_4)
\ , \\
\vecD_{00004}&=
\frac{i}{32\pi^2}
\int^1_0 \rd X_4
e^{i{n}\cdot{\mathcal{P}}_4 L}
\frac{\mathcal{M}_4}{|\mathbf{n}|}
K_1(|\mathbf{n}|L\mathcal{M}_4)
\ , \\
\vecD_{00ijk}&=
\frac{1}{32\pi^2}\int^1_0 \rd X_4
X^i_4 X^j_4 X^k_4
e^{i{n}\cdot{\mathcal{P}}_4 L}
\frac{|\mathbf{n}|L}{\mathcal{M}_4}
K_1(|\mathbf{n}|L\mathcal{M}_4)
\ , \\
\vecD_{00ij4}&=
-
\frac{iL}{32\pi^2}
\int^1_0 \rd X_4 X^i_4 X^j_4
e^{i{n}\cdot{\mathcal{P}}_4 L}
K_0(|\mathbf{n}|L\mathcal{M}_4)
\ , \\
\vecD_{00i44}&=
-\frac{1}{32\pi^2}
\int^1_0 \rd X_4 X^i_4
e^{i{n}\cdot{\mathcal{P}}_4 L}
\frac{\mathcal{M}_4 L}{|\mathbf{n}|}
K_1(|\mathbf{n}|L\mathcal{M}_4)
\ , \\
\vecD_{00444}&=
\frac{i}{32\pi^2}
\int^1_0 \rd X_4
e^{i{n}\cdot{\mathcal{P}}_4 L}
\frac{(\mathcal{M}_4)^2L}{|\mathbf{n}|^2}
K_2(|\mathbf{n}|L\mathcal{M}_4)
\ , \\
\vecD_{ijklr}&=
-\frac{1}{32\pi^2}\int^1_0 \rd X_4
X^i_4 X^j_4 X^k_4 X^l_4 X^r_4
e^{i{n}\cdot{\mathcal{P}}_4 L}
\frac{|\mathbf{n}|^2L^2}{(\mathcal{M}_4)^2}
K_2(|\mathbf{n}|L\mathcal{M}_4)
\ , \\
\vecD_{ijkl4}&=
\frac{i}{32\pi^2}\int^1_0 \rd X_4
X^i_4 X^j_4 X^k_4 X^l_4
e^{i{n}\cdot{\mathcal{P}}_4 L}
\frac{|\mathbf{n}|L^2}{\mathcal{M}_4}
K_1(|\mathbf{n}|L\mathcal{M}_4)
\ , \\
\vecD_{ijk44}&=
\frac{L^2}{32\pi^2}\int^1_0 \rd X_4
X^i_4 X^j_4 X^k_4
e^{i{n}\cdot{\mathcal{P}}_4 L}
K_0(|\mathbf{n}|L\mathcal{M}_4)
\ , \\
\vecD_{ij444}&=
-\frac{i}{32\pi^2}\int^1_0 \rd X_4
X^i_4 X^j_4
e^{i{n}\cdot{\mathcal{P}}_4 L}
\frac{\mathcal{M}_4L^2}{|\mathbf{n}|}
K_1(|\mathbf{n}|L\mathcal{M}_4)
\ , \\
\vecD_{i4444}&=
-\frac{1}{32\pi^2}\int^1_0 \rd X_4
X^i_4
e^{i{n}\cdot{\mathcal{P}}_4 L}
\frac{(\mathcal{M}_4)^2L^2}{|\mathbf{n}|^2}
K_2(|\mathbf{n}|L\mathcal{M}_4)
\ , \\
\vecD_{44444}&=
\frac{i}{32\pi^2}\int^1_0 \rd X_4
e^{i{n}\cdot{\mathcal{P}}_4 L}
\frac{(\mathcal{M}_4)^3L^2}{|\mathbf{n}|^3}
K_3(|\mathbf{n}|L\mathcal{M}_4)
\ .
\end{align}

It is straightforward to derive the relevant expressions for the $N$-point tensor integrals of higher ranks by making use of Eq.~\eqref{eq:coe.num}.

\section{Collection of tensor integrals in the CM frame}\label{app.coeff.cm}

In the CM frame, the tensor decomposition of the $N$-point integrals can be explicitly written down, in view of Eq.~\eqref{eq:CM.decom} and Eq.~\eqref{eq:CM.num}. In what follows, results shown in the preceding appendix will be reformulated accordingly.

\subsection{One-point integrals}
The one-point tensor integrals up to rank $4$ take the form of
\begin{align}
\widetilde{A}^{\mu\nu}
&=
g^{\mu\nu}\widetilde{A}_{00}
+ 
h^{\mu\nu}\widetilde{A}_{11}
\ ,\\
\widetilde{A}^{\mu\nu\rho\sigma}
&=
\{ g g \}^{\mu\nu\rho\sigma}
\widetilde{A}_{0000}
+
\{ g h \}^{\mu\nu\rho\sigma}
\widetilde{A}_{0011}
+
\{ h h \}^{\mu\nu\rho\sigma}
\widetilde{A}_{1111}
\ ,
\end{align}
with the coefficients
\begin{align}
&\widetilde{A}_{00}=\sum_{\mathbf{n}\neq 0}\vecA_{00}\ ,\quad 
\widetilde{A}_{11}=\frac{1}{d_s}\sum_{\mathbf{n}\neq 0}n^2\vecA_{11} \ , \quad 
\widetilde{A}_{0000}=\sum_{\mathbf{n}\neq 0}\vecA_{0000} \ , \notag \\
&\widetilde{A}_{0011}=\frac{1}{d_s}\sum_{\mathbf{n}\neq 0}n^2\vecA_{0011} \ ,\quad 
\widetilde{A}_{1111}=\frac{1}{(d_s+2)d_s}\sum_{\mathbf{n}\neq 0}n^4\vecA_{1111} \ ,
\end{align}
where explicit expressions of the $\vecA$ functions are shown in the previous appendix.

\subsection{Two-point integrals}
The two-point tensor integrals up to rank $4$ have the following form
\begin{align}
\widetilde{B}^{\mu}
&=
p^\mu_1 
\widetilde{B}_{1}
\ , \\
\widetilde{B}^{\mu\nu}&=
g^{\mu\nu}
\widetilde{B}_{00}
+
p^\mu_1 p^\nu_1 
\widetilde{B}_{11}
+
h^{\mu\nu} 
\widetilde{B}_{22} 
\ , \\
\widetilde{B}^{\mu\nu\rho}&=
\{ g p_1 \}^{\mu\nu\rho}
\widetilde{B}_{001}
+
p^\mu_1 p^\nu_1 p^\rho_1 
\widetilde{B}_{111}
+
\{ p_1 h \}^{\mu\nu\rho}
\widetilde{B}_{122}
\ , \\
\widetilde{B}^{\mu\nu\rho\sigma}&=\{g g\}^{\mu\nu\rho\sigma}\widetilde{B}_{0000}+\{g p_1 p_1\}^{\mu\nu\rho\sigma}\widetilde{B}_{0011}+\{g h\}^{\mu\nu\rho\sigma}\widetilde{B}_{0022}
+p^\mu_1 p^\nu_1 p^\rho_1 p^\sigma_1\widetilde{B}_{1111}
\notag \\
&
+\{p_1 p_1 h\}^{\mu\nu\rho\sigma}\widetilde{B}_{1122}
+\{h h\}^{\mu\nu\rho\sigma}\widetilde{B}_{2222}
\ ,
\end{align}
with the coefficients
\begin{align}
&\widetilde{B}_1=\sum_{\mathbf{n}\neq 0}\vecB_{1}
\ , \quad 
\widetilde{B}_{00}=\sum_{\mathbf{n}\neq0}\vecB_{00} 
\ , \quad 
\widetilde{B}_{11}=\sum_{\mathbf{n}\neq0}\vecB_{11} 
\ , \quad 
\widetilde{B}_{22}=\frac{1}{d_s}\sum_{\mathbf{n}\neq0}n^2\vecB_{22} 
\ , \notag \\
&\widetilde{B}_{001}=\sum_{\mathbf{n}\neq0}\vecB_{001} \ , 
\quad 
\widetilde{B}_{111}=\sum_{\mathbf{n}\neq0}\vecB_{111} \ ,
\quad 
\widetilde{B}_{122}=\frac{1}{d_s}\sum_{\mathbf{n}\neq0}n^2\vecB_{122} \ , \quad 
\widetilde{B}_{0000}=\sum_{\mathbf{n}\neq0}\vecB{0000} 
\ , \notag  \\
&\widetilde{B}_{0011}=\sum_{\mathbf{n}\neq0}\vecB_{0011} 
\ , \quad 
\widetilde{B}_{0022}=\frac{1}{d_s}\sum_{\mathbf{n}\neq0}n^2\vecB_{0022} 
\ , \quad 
\widetilde{B}_{1111}=\sum_{\mathbf{n}\neq0}\vecB_{1111} 
\ , \quad 
\widetilde{B}_{1122}=\frac{1}{d_s}\sum_{\mathbf{n}\neq0}n^2\vecB_{1122}
\ ,  \notag \\
&\widetilde{B}_{2222}=\frac{1}{(d_s+2)d_s}\sum_{\mathbf{n}\neq0}n^4\vecB_{2222}
\ ,
\end{align}
where explicit expressions of the $\vecB$ functions are shown in the previous appendix.

\subsection{Three-point integrals}
The three-point tensor integrals are written down up to rank $4$, which are
\begin{align}
\widetilde{C}^{\mu}
&=\sum^2_{i=1} p_i^\mu 
\widetilde{C}_{i}
\ ,  \\
\widetilde{C}^{\mu\nu}
&=g^{\mu\nu} 
\widetilde{C}_{00}
+
\sum^2_{i,j=1}p^\mu_i p^\nu_j 
\widetilde{C}_{ij}
+
h^{\mu\nu} 
\widetilde{C}_{33}
\ , \\
\widetilde{C}^{\mu\nu\rho}
&=\sum^2_{i=1}
\{ g p \}^{\mu\nu\rho}_{i}
\widetilde{C}_{00i}
+
\sum^2_{i,j,k=1}
p^\mu_i p^\nu_j p^\rho_k
\widetilde{C}_{ijk}
+
\sum^2_{i=1}
\{ p h \}^{\mu\nu\rho}_{i}
\widetilde{C}_{i33}
\ , \\
\widetilde{C}^{\mu\nu\rho\sigma}
&=
\{ g g \}^{\mu\nu\rho\sigma}
\widetilde{C}_{0000}
+
\sum^2_{i,j=1}\{ g p p \}^{\mu\nu\rho\sigma}_{ij}
\widetilde{C}_{00ij}
+
\{ g h \}^{\mu\nu\rho\sigma}
\widetilde{C}_{0033}
+
\sum^2_{i,j,k,l=1}
p^\mu_i p^\nu_j p^\rho_k p^\sigma_l
\widetilde{C}_{ijkl}
\notag \\
&
+
\sum^2_{i,j=1}\{ p p h \}^{\mu\nu\rho\sigma}_{ij}
\widetilde{C}_{ij33}
+
\{ h h \}^{\mu\nu\rho\sigma}
\widetilde{C}_{3333}
\ ,
\end{align}
with the coefficients
\begin{align}
&\widetilde{C}_{i}=\sum_{\mathbf{n}\neq0}\vecC_{i}  
\ , \quad 
\widetilde{C}_{00}=\sum_{\mathbf{n}\neq0}\vecC_{00} \ , \quad 
\widetilde{C}_{ij}=\sum_{\mathbf{n}\neq0}\vecC_{ij} \ , \quad 
\widetilde{C}_{33}=\frac{1}{d_s}\sum_{\mathbf{n}\neq0}n^2\vecC_{33} \ , \notag \\
&\widetilde{C}_{00i}=\sum_{\mathbf{n}\neq0} \vecC_{00i} \ , \quad 
\widetilde{C}_{ijk}=\sum_{\mathbf{n}\neq0} \vecC_{ijk} \ , \quad 
\widetilde{C}_{i33}=\frac{1}{d_s}\sum_{\mathbf{n}\neq0} n^2\vecC_{i33} \ , \notag \\
&\widetilde{C}_{0000}=\sum_{\mathbf{n}\neq0}\vecC_{0000} \ , \quad 
\widetilde{C}_{00ij}=\sum_{\mathbf{n}\neq0}\vecC_{00ij} \ , \quad \quad 
\widetilde{C}_{0033}=\frac{1}{d_s}\sum_{\mathbf{n}\neq0}n^2\vecC_{0033} \ , \notag \\
&\widetilde{C}_{ijkl}=\sum_{\mathbf{n}\neq0}\vecC_{ijkl} \ , \quad 
\widetilde{C}_{ij33}=\frac{1}{d_s}\sum_{\mathbf{n}\neq0}n^2\vecC_{ij33} \ , \quad 
\widetilde{C}_{3333}=\frac{1}{(d_s+2)d_s}\sum_{\mathbf{n}\neq0}n^4\vecC_{3333} \ ,
\end{align}
where explicit expressions of the $\vecC$ functions are shown in the previous appendix.
 
\subsection{Four-point integrals}
The four-point tensor integrals up to rank $5$ are given by
\begin{align}
\widetilde{D}^{\mu}
&=
\sum^3_{i=1} p^\mu_i 
\widetilde{D}_{i}
\ , \\
\widetilde{D}^{\mu\nu}
&=
g^{\mu\nu} 
\widetilde{D}_{00}
+
\sum^3_{i,j=1}
p^\mu_i p^\nu_j 
\widetilde{D}_{ij}
+
h^{\mu\nu}
\widetilde{D}_{44}
\ , \\
\widetilde{D}^{\mu\nu\rho}
&=
\sum^3_{i=1}\{ g p \}^{\mu\nu\rho}_i
\widetilde{D}_{00i}
+
\sum^3_{i,j,k=1}p^\mu_i p^\nu_j p^\rho_k \widetilde{D}_{ijk}
+
\sum^3_{i=1}
\{ p h \}^{\mu\nu\rho}_i
\widetilde{D}_{i44}
\ , \\
\widetilde{D}^{\mu\nu\rho\sigma}
&=
\{ g g \}^{\mu\nu\rho\sigma}
\widetilde{D}_{0000}
+
\sum^3_{i,j=1}\{ g p p \}^{\mu\nu\rho\sigma}_{ij}
\widetilde{D}_{00ij}
+\{ g h \}^{\mu\nu\rho\sigma}\widetilde{D}_{0044}
+
\sum^3_{i,j,k,l=1}p^\mu_i p^\nu_j p^\rho_k p^\sigma_l 
\widetilde{D}_{ijkl}
\notag \\
&
+
\sum^3_{i,j=1}\{ p p h \}^{\mu\nu\rho\sigma}_{ij}
\widetilde{D}_{ij44}
+
\{ h h \}^{\mu\nu\rho\sigma}
\widetilde{D}_{4444}
\ ,  \\
\widetilde{D}^{\mu\nu\rho\sigma\alpha}
&=
\sum^3_{i=1}
\{ g g p \}^{\mu\nu\rho\sigma\alpha}_{i}
\widetilde{D}_{0000i}
+
\sum^3_{i,j,k=1}
\{ g p p p \}^{\mu\nu\rho\sigma\alpha}_{ijk}
\widetilde{D}_{00ijk}
+
\sum^3_{i=1}
\{ g p h \}^{\mu\nu\rho\sigma\alpha}_{i}
\widetilde{D}_{00i44}
\notag \\
&
+
\sum^3_{i,j,k,l,r=1}
p^\mu_i p^\nu_j p^\rho_k p^\sigma_l p^\alpha_r 
\widetilde{D}_{ijklr}
+
\sum^3_{i,j,k=1}
\{ p p p h \}^{\mu\nu\rho\sigma\alpha}_{ijk}
\widetilde{D}_{ijk44}
+
\sum^3_{i=1}
\{ p h h \}^{\mu\nu\rho\sigma\alpha}_i
\widetilde{D}_{i4444}
\ ,
\end{align}
with the coefficients
\begin{align}
&\widetilde{D}_{i}=\sum_{\mathbf{n}\neq0}\vecD_{i} 
\ , \quad 
\widetilde{D}_{00}=\sum_{\mathbf{n}\neq0}\vecD_{00}
\ , \quad 
\widetilde{D}_{ij}=\sum_{\mathbf{n}\neq0}\vecD_{ij}
\ , \quad 
\widetilde{D}_{44}=\frac{1}{d_s}\sum_{\mathbf{n}\neq0}n^2\vecD_{44}
\  , \notag \\
&\widetilde{D}_{00i}=\sum_{\mathbf{n}\neq0}\vecD_{00i}  \ , \quad 
\widetilde{D}_{ijk}=\sum_{\mathbf{n}\neq0}\vecD_{ijk}  \ , \quad 
\widetilde{D}_{i44}=\frac{1}{d_s}\sum_{\mathbf{n}\neq0}n^2\vecD_{i44} 
\ , \notag \\
&\widetilde{D}_{0000}=\sum_{\mathbf{n}\neq0}\vecD_{0000} \ , \quad 
\widetilde{D}_{00ij}=\sum_{\mathbf{n}\neq0}\vecD_{00ij} \ , \quad 
\widetilde{D}_{0044}=\frac{1}{d_s}\sum_{\mathbf{n}\neq0}n^2\vecD_{0044} \ , \notag \\
&\widetilde{D}_{ijkl}=\sum_{\mathbf{n}\neq0}\vecD_{ijkl} \ , \quad 
\widetilde{D}_{ij44}=\frac{1}{d_s}\sum_{\mathbf{n}\neq0}n^2\vecD_{ij44} \ , \quad
\widetilde{D}_{4444}=\frac{1}{(d_s+2)d_s}\sum_{\mathbf{n}\neq0}n^4\vecD_{4444} \ , \notag \\
&\widetilde{D}_{0000i}=\sum_{\mathbf{n}\neq0}\vecD_{0000i} \ , \quad 
\widetilde{D}_{00ijk}=\sum_{\mathbf{n}\neq0}\vecD_{00ijk} \ , \quad 
\widetilde{D}_{00i44}=\frac{1}{d_s}\sum_{\mathbf{n}\neq0}n^2\vecD_{00i44}
\ , \notag \\
&\widetilde{D}_{ijklr}=\sum_{\mathbf{n}\neq0}\vecD_{ijklr} \ , \quad 
\widetilde{D}_{ijk44}=\frac{1}{d_s}\sum_{\mathbf{n}\neq0}n^2\vecD_{ijk44} \ , \quad 
\widetilde{D}_{i4444}=\frac{1}{(d_s+2)d_s}\sum_{\mathbf{n}\neq0}n^4\vecD_{i4444} \ ,
\end{align}
where explicit expressions of the $\vecD$ functions are shown in the previous appendix.

\section{One-loop self-energies of the nucleon \label{app.se}}
The dressed propagator of the nucleon is given by
\bea
iS_N(p)=\frac{i}{\slashed{p}-m-\Sigma(\slashed{p})}\ ,
\eea
where $\Sigma(\slashed{p})$ and $m$ are the nucleon self-energy and bare mass, respectively. The physical nucleon mass $m_N$ is defined as the pole of the dressed propagator,
\bea
m_N=m+\Sigma(m_N)\ .
\eea
Feynman diagrams displayed in Fig.~\ref{fig:m} contribute to the self-energies at one-loop level. 

At finite volume, the corresponding FVC effects have been calculated in, e.g., Ref.~\cite{Alvarez-Ruso:2013fza}. Here we list the explicit expressions for the sake of easy comparison. For diagram (a), the self-energy reads
\begin{align}
\Sigma_{N3}(M_{\pi}^2,L)
&=3\left[\frac{g_A}{8F_{\pi}\pi}\right]^2
\sum_{\mathbf{n}\neq0}
\int^1_0 {\rm d}z 2 m_N
\bigg[
((1-z)^3 m_N^2+(3-z)\mathcal{M}_N^2) K_0(|\mathbf{n}|L\sqrt{\mathcal{M}_N^2})
\notag \\
&
+
(4z-6)\frac{\sqrt{\mathcal{M}_N^2}}{L|\mathbf{n}|} K_1(|\mathbf{n}|L\sqrt{\mathcal{M}_N^2})
\bigg] \ ,\label{eq:ruso.mN}
\end{align}
with $\mathcal{M}_N^2=z M_{\pi}^2-z(1-z)p^2+(1-z) m_{N}^2$. Here $g_A$ and $F_\pi$ are the nucleon axial coupling and the pion decay constants, respectively. For diagram (b), the self-energy is
\begin{align}
\Sigma_{N\Delta3}(M_{\pi}^2,L)
&=\frac{4}{3}\left[\frac{h_A}{8F_{\pi}\pi m_{\Delta}}\right]^2
\sum_{\mathbf{n}\neq0}
\int^1_0 {\rm d}z(z m_N+m_{\Delta})
2m_N^2
\bigg[ 
-\frac{\sqrt{\mathcal{M}_{\Delta}^2}}{L|\mathbf{n}|} K_1(|\mathbf{n}|L\sqrt{\mathcal{M}_{\Delta}^2})
\notag \\
&
+\mathcal{M}_{\Delta}^2 K_0(|\mathbf{n}|L\sqrt{\mathcal{M}_{\Delta}^2})
\bigg] \ ,
\end{align}
with $\mathcal{M}_{\Delta}^2=z M_{\pi}^2-z(1-z)p^2+(1-z) m_{\Delta}^2$ and $h_A$ being the axial coupling constant of the $N$-$\Delta$ transition. Mention that the above expressions are obtained in the rest frame of the nucleon.

In our case, diagram (a) and diagram (b) yield
\begin{align}
\Sigma_a(\slashed{p},\slashed{n})
&=\frac{g_A^2}{4F^2_{\pi}}
\sum_{{\bf n}\neq 0}\frac{1}{i}\int\frac{{\rm d}^d k}{(2\pi)^d}e^{-il_k\cdot k}
\frac{\slashed{k}[(\slashed{k}+\slashed{p})-m_N]\slashed{k}}{[(k+p)^2-m_N^2][k^2-M_{\pi}^2]}
\ , \notag \\
\Sigma_b(\slashed{p},\slashed{n})
&=\frac{h_A^2}{F^2_{\pi}}
\sum_{{\bf n}\neq 0}\frac{1}{i}\int\frac{{\rm d}^d k}{(2\pi)^d}e^{-il_k\cdot k}
\frac{(\slashed{k}+\slashed{p})+m_{\Delta}}{[(k+p)^2-m_{\Delta}^2][k^2-M_{\pi}^2]}
\notag \\
&
\times
\bigg\{
\frac{k^2}{d-1}
+\frac{(d-2)(k\cdot(k+p))^2}{(d-1)m_{\Delta}^2}
-k^2
\bigg\}
\ .
\end{align}
By making use of the FVC tensor integrals defined in section~\ref{sec.decom}, the above expressions of self-energies become 
\begin{align}
\Sigma_a(\slashed{p},\slashed{n})&=\frac{3g_A^2m_{N}}{4F^2_{\pi}}\sum_{\mathbf{n}\neq 0}\bigg\{
s \vecB_0+2s \vecB_1+d \vecB_{00}+s \vecB_{11}+n^2\vecB_{22}
-2 n\cdot p \left[\vecB_2+\vecB_{12}\right]
\bigg\}\ ,\notag\\
&+\frac{3g_A^2}{4F^2_{\pi}}\sum_{\mathbf{n}\neq 0}\bigg\{
s \vecB_1+2s \vecB_{11}+2d \vecB_{00}+
(d+2) \vecB_{001}
+s \vecB_{111}
\notag \\
& \hspace{2cm}
+n^2\left[2\vecB_{22}+\vecB_{122}\right]
-2n\cdot p\left[\vecB_2+2\vecB_{12}+\vecB_{112}\right]
\bigg\}\slashed{p}\ ,\notag\\
&+\frac{3g_A^2}{4F^2_{\pi}}\sum_{\mathbf{n}\neq 0}\bigg\{
s \vecB_2
-(d+2)\vecB_{002}
-s\vecB_{112}
-n^2\vecB_{222}
+2n\cdot p \vecB_{122}
\bigg\}\slashed{n}\ ,\label{eq.se.a.long}
\end{align}
and
\begin{align}
\Sigma_b(\slashed{p},\slashed{n})
&=
\frac{2(d-2)h_A^2}{(d-1)F_{\pi}^2 m_{\Delta}}
\sum_{\mathbf{n}\neq0}
\bigg\{
m_{\Delta}^2 s \vecB_0
+2m_{\Delta}^2 s \vecB_{1}
+(d m_{\Delta}^2-s) \vecB_{00}
+s(m_{\Delta}^2-s)\vecB_{11}
\notag \\
&
-2s(2+d) \vecB_{001}
-2s^2\vecB_{111}
-d(2+d)\vecB_{0000}
-2s(2+d)\vecB_{0011}
-s^2 \vecB_{1111}
\notag \\
&
+n^2\bigg[ m_{\Delta}^2\vecB_{22}
-2s \vecB_{122}
-2(2+d)\vecB_{0022}
-2s\vecB_{1122}
-n^2\vecB_{2222}\bigg] 
\notag \\
&
-
2 n\cdot p
\bigg[
m_{\Delta}^2 \vecB_{2}
-(s-m_{\Delta}^2) \vecB_{12}
-(2+d)(\vecB_{002}+2\vecB_{0012})
-s(3\vecB_{112}+2\vecB_{1112})
\notag \\
&
-n^2(\vecB_{222}+2\vecB_{1222})
\bigg] 
-(n\cdot p)^2 \bigg[\vecB_{22}+4\vecB_{122}+4\vecB_{1122}\bigg] 
\bigg\}
\notag \\
&
+
\frac{2(d-2)h_A^2}{(d-1)F_{\pi}^2 m_{\Delta}^2}
\sum_{\mathbf{n}\neq0}
\bigg\{
-m_{\Delta}^2 s \vecB_{1}
-2m_{\Delta}^2 \vecB_{00}
-2m_{\Delta}^2 s \vecB_{11}
-(m_{\Delta}^2(2+d)-3s)\vecB_{001}
\notag \\
&
-s(m_{\Delta}^2-s)\vecB_{111}
+2(2+d)\vecB_{0000}
+2s(5+d)\vecB_{0011}
+2s^2\vecB_{1111}
+(d+2)(d+4)\vecB_{00001}
\notag \\
&
+2s(4+d)\vecB_{00111}
+s^2\vecB_{11111}
-n^2\bigg[
m_{\Delta}^2\vecB_{122}
-2\vecB_{0022}
-2s\vecB_{1122}
-2(4+d)\vecB_{00122}
\notag \\
&
-2s\vecB_{11122}
-n^2\vecB_{12222}
\bigg]
+
2n\cdot p
\bigg[
m_{\Delta}^2\vecB_{12}
-\vecB_{002}
-(s-m_{\Delta}^2)\vecB_{112}
-(6+d)\vecB_{0012}
\notag \\
&
-2(4+d)\vecB_{00112}
-s(3\vecB_{1112}+2\vecB_{11112})
-n^2(\vecB_{1222}+2\vecB_{11222})
\bigg] 
\notag \\
&
+
(n\cdot p)^2
\bigg[\vecB_{122}+4\vecB_{1122}+4\vecB_{11122}\bigg] 
\bigg\}\slashed{p}
\notag \\
&
+
\frac{2(d-2)h_A^2}{(d-1)F_{\pi}^2 m_{\Delta}^2}
\sum_{\mathbf{n}\neq0}
\bigg\{
m_{\Delta}^2 s \vecB_{2}
+2m_{\Delta}^2 s \vecB_{12}
+(m_{\Delta}^2 (2+d)-s)\vecB_{002}
+s(m_{\Delta}^2-s)\vecB_{112}
\notag \\
&
-2s(4+d)\vecB_{0012}
-2s^2\vecB_{1112}
-(d+2)(d+4)\vecB_{00002}
-2s(d+4)\vecB_{00112}
-s^2\vecB_{11112}
\notag \\
&
+
n^2\bigg[
m_{\Delta}^2\vecB_{222}
-2s\vecB_{1222}
-2(d+4)\vecB_{00222}
-2s\vecB_{11222}
-n^2\vecB_{22222}
\bigg] 
\notag \\
&
-2n\cdot p 
\bigg[
m_{\Delta}^2\vecB_{22}
-(s-m_{\Delta}^2)\vecB_{122}
-(4+d)(\vecB_{0022}+2\vecB_{00122})
-s(3\vecB_{1122}+2\vecB_{11122})
\notag \\
&
-
n^2(\vecB_{2222}+2\vecB_{12222})
\bigg] 
-
(n\cdot p)^2 \bigg[\vecB_{222}+4\vecB_{1222}+4\vecB_{11222}\bigg]
\bigg\}\slashed{n}  \ .
\label{eq.se.b.long}
\end{align}


\end{document}